\documentclass[final,1p,times]{elsarticle}

\usepackage{amssymb}
\usepackage{graphicx}
\usepackage{amsmath}
\usepackage{float}

\usepackage{color}

\biboptions{numbers}

\newcounter{bla}

\journal{Computer Physics Communications}

\begin{document}


\def\Msol{M_{\odot}} 
\def\Lsol{L_{\odot}}
\def\Rsol{R_{\odot}} 
\def\Rast{R_\ast} 
\def\Mast{M_\ast}
\def\Teff{T_{\rm eff}}
\def\cs{c_{\rm s}}
\def\uf{u_\varphi}
\def\pop#1#2{\frac{\partial#1}{\partial#2}}
\def\rip{r_\mathrm{i+1}}
\def\rim{r_\mathrm{i-1}}
\def\rih{r_\mathrm{i+\frac{1}{2}}}
\def\E{1\!\!1}

\begin{frontmatter}



\title{1+1D implicit disk computations}


\author[a]{Florian Ragossnig\corref{author}}
\author[a]{Ernst Dorfi}
\author[a]{Bernhard Ratschiner}
\author[a]{Lukas Gehrig}
\author[a]{Daniel Steiner}
\author[a]{Alexander St\"okl}
\author[a]{Colin P. Johnstone}

\cortext[author] {Corresponding author.\\\textit{E-mail address:} florian.ragossnig@univie.ac.at}
\address[a]{University of Vienna, Department of Astrophysics, T\"urkenschanzstr. 17, 1180 Vienna, Austria}

\begin{abstract}

We present an implicit numerical method to solve the time-dependent equations of radiation hydrodynamics (RHD) in axial symmetry assuming hydrostatic equilibrium perpendicular to the equatorial plane (1+1D) of a gaseous disk. The equations are formulated in conservative form on an adaptive grid and the corresponding fluxes are calculated by a spacial second order advection scheme. Self-gravity of the disk is included by solving the Possion equation. We test the resulting numerical method through comparison with a simplified analytical solution as well as through the long term viscous evolution of protoplanetary disk when due to viscosity matter is transported towards the central host star and the disk depletes. The importance of the inner boundary conditions on the structural behaviour of disks is demonstrated with several examples.


\end{abstract}


\end{frontmatter}

\section{Introduction}
\label{s.intro}
During the gravitational collapse of tenuous interstellar gas towards a condensed object like a protostar, the initial angular momentum has to be reduced by several orders of magnitude to overcome the centrifugal barrier \cite{Mestel1965, Schatzman1962}. A solution to this angular momentum problem is the formation of an accretion disk around the central star where turbulence-induced viscosity and/or magnetic fields provide a mechanism to transport angular momentum outwards accompanied by a mass transport towards the interior regions.

We point out that accretion disks do not only occour as protoplanetary disks (PPDs) but also around other astrophysical objects like black holes and white dwarfs. Such accretion disks are usually characterized by large disk masses, high surface irradiation and/or the disks are cut off due to the Roche lobe of a companion star feeding the disk \citep[e.g.][]{Hameury1998}. Since non-PPDs are either massive (disks around black holes) or experience a mass flux over the outer boundary (dwarf-novae), they are unlikely to be dynamically stable. Although, we are confident that our model can be adopted to such disks, we want to emphasize that our current approach of an isothermal, vertical disk-structure is approximatevely valid for PPDs. However, the vertical structure of disks around x-ray binaries and white dwarfs requires a more careful treatment of the vertical radiative transport, in particular to get the correct, opacity-dependent photospheric temperature stratification. We accentuate that this work focuses on protoplanetary accretion disks with disk masses of less than 10\% of the stellar mass and an outer boundary located in the interstellar medium ISM (cf. Sec.~\ref{ss.bc}).

 In the simplest situation, we can consider a disk with negligible mass around a central object with mass $\Mast$ in centrifugal balance and obtain a Keplerian angular velocity at a radius $r$
\begin{equation} \label{e.kepler}
     \Omega = \sqrt{\frac{\mathrm{G} \Mast}{r^3}} \:,
\end{equation}
with an orbital velocity of $\uf = \Omega\,r \propto r^{-1/2}$. Hence, assuming an orbital ratio of $100$, the corresponding orbital periods vary according to Kepler's 3$^{rd}$ law by a factor of $1000$. This basic physical requirement immediately hinder all numerical simulations of disk-like structures. Taking $\Delta x$ as an orbital distance, any explicit numerical computation is restricted by the Courant, Friedrichs and Lewy (CFL) condition \citep{Courant1928} 
\begin{equation} \label{e.CFL}
   \delta t \le \min_{\rm{all\, cells}}
     \frac{\Delta x}{\vert\uf\vert + \cs} 
     \simeq \min_{\rm IB} \frac{\Delta x}{\vert\uf\vert} \:.
\end{equation}
For stability, the smallest timesteps $\delta t$ at the inner boundary (IB) must be taken. Although it is possible to increase the timestep of an explicit scheme beyond the CFL limit by special methods (e.g. FARGO, local timestepping, \citep{Masset2000, Liska2018}), long term evolution studies of PPDs with an inner boundary within a few stellar radii away from the star, are only possible by utilizing implicit methods. All disks are dominated by centrifugal forces, and thus the sound velocity of the gas, $\cs$, is small compared to the orbital velocity, $\uf$ (i.e.~$\cs\ll \uf$). 
 
Almost all astrophysical disk simulations are carried out by explicit numerical schemes and as a result of the CFL condition, the size of the computational domain towards the inner regions of the disk is limited \cite[e.g.][]{Vorobyov2018}. For matter orbiting the host star, Keplerian rotation leads to very large velocity gradients which results in very small timesteps, typically of the order of one orbital period of the inner region divided by the number of azimuthal grid points. Taking into account infrared-observations \cite[e.g.][]{Hernandez2008} of protoplanetary disks, the estimated lifespan of a disk can be estimated to be up to $10^7\,$years. Therefore, the application of explicit schemes is limited, if one wants to investigate the long term evolution of disks. To overcome this restriction, we utilize an implicit time integration scheme which allows us to extend the inner boundary to the surface of the host star and is not restricted by timestep length \cite[e.g.][]{Dorfi1998}. For physical reasons we usually set the inner boundary to the co-rotation radius, which is where the rotational velocity of the star matches the angular velocity of the disk gas (see Sec.~\ref{ss.bc}) and is typically at few stellar radii. We emphasize the importance of the inner boundary condition as the solution of disk evolution is strongly coupled to changes of flux and temperature within the innermost disk. Recent studies of viscous disk models support this argument that an appropriate description of the innermost region is a significant factor for global disk evolution \cite{Vorobyov2019} and cannot be approximated by specially designed inner sink cells.

Usually, boundary conditions for partial differential equations (PDE) determine the solution out of a class of functions and are therefore more important and more complicated to implement compared to ordinary differential equations. In particular, computing disk-like structures can strongly depend on the location of the boundaries because neither the inner nor the outer boundary is clearly determined by geometrical conditions. Consequently, the disk mass depends entirely on the choice of the inner and outer boundary condition.

Furthermore, the time step in implicit schemes does not depend on cell size but is entirely determined by the accuracy and convergence of the utilized method. If short timescale phenomena are present, implicit schemes bear no advantage over explicit schemes. Utilizing an implicit method and implementing axial symmetry can lead to time steps of several orbital periods and allows us to study the evolution of interplanetary disks throughout the entire disk lifetime.

Although our method adopts axial symmetry and hence no azimuthal resolution can be obtained, recent {\sf ALMA} (Atacama Large Millimeter Array) observations show that 75\% of the disks around young T~Tauri stars are axially symmetric protoplanetary disks, at least within the resolution limit of the instrument (approximately $5\,$AU) \cite{Andrews2018}. Disks showing a spiral-like structures or small anti-symmetric features either involve a companion star or are most likely disturbed by an undetected planet \cite{Perez2018, Isella2018}. Furthermore, recent simulations show that especially less massive disks tend towards axial symmetry shortly after the initial collapse and remain about 80\% of their life in an axisymmetric state \cite[see e.g.][]{Vorobyov2019}. Encouraged by such observations and non-axisymmetric simulations, we adopt a cylindrical geometry 
    to study the long term evolution of protoplanetary disks.

This paper describes recent developments of our initially spherical symmetric {\sf TAPIR} code (The Adaptive Implicit RHD Code) \cite{Dorfi1987}, now modified for cylindrical configurations and in particular the geometrical changes necessary to ensure also a conservative formulation of the equations of radiation hydrodynamics on an adaptive grid. Furthermore, self-gravity of the disk material has also been included by solving the corresponding Poisson equation for an axisymmetric density distribution. In Section~\ref{s.phys_equations}, we describe the physical equations implemented in our model. In Section~\ref{s.num}, we present the numerical methods. Section~\ref{s.example} is devoted to numerical examples and accuracy tests to validate our computations. Finally, in Section~\ref{s.conclusion}, we discuss some simulations concerning their astrophysical perspectives.

\section{Physical equations}
\label{s.phys_equations}
We adopt a cylindrical coordinate system with $(r,\varphi,z)$ and assume hydrostatic equilibrium in the $z$-direction and therefore no vertical velocities. The velocity vector is therefore ${\bf u}=(u_r(r,t), \uf(r,t),0)$. Also, we neglect vertical variations for the velocity and the internal energy resulting in a constant sound velocity $\cs$ perpendicular to the equatorial plane. To eliminate further vertical dependences (e.g.~in the equation of motion) we adopt the thin disk approximation ($z \ll r$), which in particular is given by $(r^2+z^2)^{-3/2} \sim r^{-3}$ and results in a local constant vertical gravitational acceleration. Since all quantities are independent on the azimuthal angle $\varphi$, we can use vertically integrated quantities. For example, the surface density $\Sigma$ is related to the mass density $\rho$ by 
\begin{equation}  \label{e.surf}
    \Sigma(r,t) = \int_{-\infty}^{\infty} \rho(r,z,t)\,dz \:.
\end{equation}
Correspondingly, the equation of continuity is given by \cite[e.g.][]{LL1987}
\begin{equation}  \label{e.cont}
  \pop{\Sigma}{t} + \frac{1}{r}\pop{}{r} \left( ru_r\Sigma \right) = 0 \:.
\end{equation}

As a result of utilizing cylindrical geometry, the equation of motion appears in a radial and an angular part. Applying the above assumptions to the radial equation of motion gives
\begin{multline} \label{e.mot_r}
    \pop{}{t} \left( \Sigma u_r \right)
        + \frac{1}{r} \pop{}{r} \left( r \, \Sigma \, u_r^2 \right) - \frac{\rho \, u_\varphi^2}{r}
        + \pop{P}{r} \\
        + \Sigma \pop{\Psi_\mathrm{tot}}{r}
        - \frac{4 \pi}{c} \kappa_{\rm R} \Sigma H_r
        + \frac{1}{r} \pop{}{r} \left( r \, Q_{rr} \right) - \frac{Q_{\varphi \varphi}}{r} = 0 \:,
\end{multline} 
where $\kappa_\mathrm{R}$ is the Rosseland-mean opacity \cite[e.g.][]{Mihalas1984}, the thermodynamic quantity $P$ represents the vertically integrated gas pressure and is retrieved from tables, $H_r$ represents the radial radiation flux component, i.e.~1\textsuperscript{st} moment of the specific intensity and $Q_{rr}$ and $Q_{\varphi \varphi}$ are the radial and angular components of the viscous pressure tensor, respectively \cite[][]{Tscharnuter1979}. The generalised form of $Q$ writes (written as unity matrix)
\begin{equation}
    Q = \mu_Q \left[ (\nabla u) - \frac{1}{3} \nabla \cdot u \E \right]
\end{equation}
The quantity $\partial\Psi_\mathrm{tot}/\partial{r} = \partial\Psi_\mathrm{s}/\partial{r} + \partial\Psi_\mathrm{d}/\partial{r} $ represents the total gravitational potential, where $\partial\Psi_\mathrm{s}/\partial{r} = G \Mast / r^2$ is the star's potential acceleration (cf Sect.~\ref{ss.pot}).

Since thermal, gravitational and radiation pressures do not contribute to the $\varphi$-component of the equation of motion, the appropriate terms can be neglected. After some simplifications, the angular component of the equation of motion becomes
\begin{equation} \label{e.mot_phi}
    \pop{}{t} ( r \Sigma u_\varphi )
        + \dfrac{1}{r} \pop{}{r} \left( r^2 \Sigma u_r u_\varphi \right) 
        + \pop{}{r} \left( r \, Q_{r \varphi} \right) + Q_{\varphi r} = 0 \:.
\end{equation}

Hydrostatic equilibirum in the $z$-direction leads to a simple vertial density structure given by \cite[e.g.][]{Armitage2013}
\begin{equation}
    \rho(r,z) = \rho_0(r) \exp{\left( -\frac{z^2}{2H_\mathrm{p}^2} \right)} \:,
\end{equation}
where $\rho_0(r)$ is the equatorial density. The last expression uses a defintion of scale height, $H_\mathrm{p}$, given by
\begin{equation}  \label{e.hp}
    H_\mathrm{p} = \frac{\cs^2}{g_\mathrm{z}} \:,
\end{equation}
and note, that $\cs$ and the vertical gravitational acceleration $g_\mathrm{z}$ depend on the radial distance $r$. Since we adopt the thin disk approximation \citep[e.g.][page 39]{Armitage2013} (valid for $z \ll r$)
\begin{equation}
    g_z = \frac{G M_{\ast} z}{r^3} \:.
\end{equation}
A more elaborated treatment involves the solution of the Poisson equation (see Sect.~\ref{ss.pot}).

The generalized form of the equation of internal energy in cylindrical geometry is
\begin{multline} \label{e.ene_gen}
    \pop{}{t} \left( \Sigma e_r \right)
        + \frac{1}{r}\pop{}{r} \left( ru_r\Sigma e_r \right) 
        + P \frac{1}{r}\pop{}{r} \left( ru_r \Sigma \right) 
        + \epsilon_Q \\
        - 4\pi\kappa_\mathrm{P} \left( J-S \right)
        + \Delta E_\mathrm{rad}
        + \frac{1}{r}\pop{}{r} \left( r u_\phi \right) = 0 \:,
\end{multline}
where $\epsilon_Q = {\bf Q} : {\bf \nabla} {\bf u}$ (where $:$ is the tensor product) denotes the viscous energy dissipation \cite[][]{Tscharnuter1979}, $\kappa_\mathrm{P}$ is the Planck opacity \cite[e.g.][]{Mihalas1984} and $\Delta E_\mathrm{rad}$ is the radiative heating/cooling term which will be discussed in Sect.~\ref{s.num} in more detail. Since the radial velocity $u_r$ is much smaller than the angular velocity $u_\varphi$, we can truncate the radiation equations to their dominant terms
\begin{equation} \label{e.nabla_H}
    \kappa_\mathrm{P} \Sigma \left( J-S \right) = -{\bf \nabla} \cdot {\bf H}
\end{equation}
and
\begin{equation} \label{e.nabla_K}
    \kappa_\mathrm{R} \Sigma {\bf H} = -{\bf \nabla} \cdot {\bf K} \sim -{\bf \nabla} \left( f_\mathrm{edd}J \right) \:,
\end{equation}
where $f_\mathrm{edd}$ represents the Eddington factor \cite[e.g.][]{Mihalas1984}. We emphasize that these dominant terms are purely spatial and will not obey any time scale. As the time-dependent terms in the radiation flux equation are usually not important compared to the pure radiation diffusion terms, they can be neglected. Implementing Eq.~\ref{e.nabla_H} and \ref{e.nabla_K} into Eq.~\ref{e.ene_gen}, we get
\begin{multline} \label{e.ene_ax}
    \pop{}{t} \left( \Sigma e_r \right)
        + \frac{1}{r}\pop{}{r} \left( ru_r\Sigma e_r \right) 
        + P \frac{1}{r}\pop{}{r} \left( ru_r \Sigma \right) 
        + \epsilon_Q \\
        + 4\pi \frac{1}{r} \pop{}{r} \left(r H_r\right)
        + \Delta E_\mathrm{rad}
        + \frac{1}{r}\pop{}{r} \left( r u_\phi \right) = 0 \:.
\end{multline}

For a consistent description of the RHD equations, we would have to formulate the equation of the radiation flux and the equation for the radiation energy as individual equations. Since the internal energy budget is dominated by irradiation and radiative cooling, we can simplify the description for the radiation temperature $J$ and the radiative flux ${\bf H}$ further, and implement both directly into the internal energy equation. A more detailed description of how this problem has been tackled is given in Sect.~\ref{ss.discrete_equations}.

\subsection{Boundary conditions}
\label{ss.bc}
As known from the theory of PDEs, boundary conditions will determine the solution to a particular problem. In the case of astrophysical disks, the locations and the physical properties of the boundaries are difficult to specify. Since our computations are restricted to axisymmetric problems, we have to avoid massive disks where local gravitational instabilities can lead to a collapse of individual gas blobs \cite{Vorobyov2018}. In our case, we only assume non-massive disks where $M_{\rm disk} \ll \Mast$ and where the gravitational potential is dominated by the central star with mass $\Mast$. Typically, non-axial structures will grow if $M_{\rm disk}$ exceeds approximately $0.1\, \Mast$ \cite[e.g.][]{Armitage2013}. The disk's gravitational potential is smaller by several orders of magnitude (cf. Sect.~\ref{ss.pot} for details). 

The exact position of the inner and outer boundary of PPDs is hard to define since the inner boundary depends on the physical properties of the star and the outer boundary is basically defined at the location of the interstellar medium (ISM). At the inner boundary, a star with a finite radius rotates at a different angular speed $\Omega_\ast$ than  the material of the circumstellar disk $\Omega(r)$. The corotation radius, $r_{\rm co}$, is defined as the radius where the stellar rotation rate matches the orbital speed of the disk, i.e. $\Omega_\ast=\Omega(r_{\rm co})$, and depends on the topology and the strength of the stellar magnetic field \citep{Hartmann2016}. Due to the centrifugal balance of the disk (cf.~Eq.~\ref{e.kepler}), $r_\mathrm{co}$ determines the gap between the star and the disk. Details about the mass flow of disk material onto the star can only be analyzed by detailed 3D-MHD simulations (\cite[e.g][]{Romanova2009}). The role of the stellar magnetic field emerges also in the so-called magnetic truncation radius $r_{\rm{mt}}$ \citep{Hartmann2016}, where studies place the inner boundary of a disk at the location where the stellar magnetic field matches the disk's magnetic field \citep[X-point, see][]{Shu2000}. In some cases, depending on the disk mass and in case of a very weak magnetic field, it is possible that the disk reaches to the surface of the star \citep{Belyaev2013}. Regardless of the definite process, the position of the inner boundary for the majority of PPDs is close to $r_{\rm co}$ and within a few stellar radii away from the host star. Although our model is capable of setting the inner boundary at any radius (also at the star's surface), the intention of this work is to present a model to investigate the long term evolution of PPDs. Thus we do not focus on detailed astrophysical applications but on the importance of the ability of a model to set the inner disk boundary close to the star. In Sect.~\ref{s.example} we show that the solutions are critically affected if a simplistic central hole is cut out of the computational domain. Hence we emphasize that the treatment of the inner boundary is crucial to obtain reasonable physical models and neglecting the innermost regions can critically distort the disk structure.

Since most PPDs do not have a companion and therefore no mass inflow over the outer disk boundary, the outer boundary is less important at distances far from the central source. Note that the exact position of the outer boundary does become more important when dealing with massive disks (e.g. around black holes) and/or the disk is getting fed by a companion \citep{Hameury1998}. Again, examples are given in Sect.~\ref{s.example}. In principle, our model allows to define the outer boundary at very large radii but to avoid numerical issues, we set the outer disk radius to a value where the denisty does not fall below $\rho_{\rm{out}} = 10^{-8}\,$g/cm$^3$ according to the standard disk model \citep{Armitage2013}. For example, if the density drops below the numerical accuracy, we can set it to values typical for the ISM near the outer disk edge. Since implicit numerical schemes are not restricted by the CFL timestep condition, the implementation of appropriate boundary conditions near to the centre as well as far outside enables large space regions to be covered and therefore opens another advantage for the global disk simulations described in the next sections.

\subsection{Viscosity}
\label{ss.vis}
Astrophysical disks are characterized by viscous forces due to turbulent motions which enable the transport of circulating material towards the central source. This accretion process reduces the mass of the disk but the physical nature of the viscosity is still under debate \cite{Balbus1991, Lin1987}. The timescale of disk dispersal by viscosity is much longer than the Keplerian orbital time which is demonstrated in Sect~\ref{ss.density_waves} and Fig.~\ref{f.armitage}.

The usual viscosity description relies on the turbulence based approach for the kinematic viscosity $\nu$ according to Shakura \& Sunyayev \cite{Shakura1973}
\begin{equation} \label{e.shak}
  \nu  = \alpha \cs H_\mathrm{P} \:,
\end{equation}
where $\alpha$ is a free parameter, $\cs$ is the sound velocity and $H_\mathrm{P}$ the vertical scale height (cf.~\ref{e.hp}). Such a formulation is based on the idea that turbulent gas blobs are moving at speed similar to the sound speed and that the largest turbulent eddies are naturally limited by the scale height of the disk. Numerical simulations of the magnetic rotational instability MRI \cite[e.g.][]{Bai2014} can be used to determine a more reliable estimate of $\alpha$ and typical values are assumed to be $\alpha\sim 0.01$. Note that this value for alpha represents the ideal MHD limit. In a realistic astrophysical scenario alpha certainly can have a more complicated radial dependency \citep[e.g.][]{Balbus1991}. In this work we do not intent to investigate such realistic astrophysical applications but describe in detail a novel method to study the evolution of PPDs. The dependency of the alpha parameter and the evolution of a simple viscous disk is discussed in more detail in Sect.~\ref{s.example}.

\subsection{Gravitational Potential}
\label{ss.pot}
Although the disk mass is small compared to the stellar mass, changes in the disk's gravitational potential can play an important role when mass is redistributed within the disk. Although we do not focus on detailed astrophysical scenarios where the disk potential becomes relevant (e.g.~so-called FU Ori outbursts) we include in our numerical method the disk's gravitational potential  to obtain a more consistent modelling. Furthermore, we mention that the majority of PPDs have masses around 1\% to 5\% of the stellar mass. Such disks are in any case gravitationally stable with respect to the Toomre parameter of $Q > 100$. Usually, protoplanetary disks become gravitationally unstable around $Q=1$, i.e.~the disk mass reaches around 10\% of the stellar mass \citep[e.g.][]{Lodato2004, Boley2006, Cossins2009}.

The gravitational potential and the gravitational acceleration towards the central star of a thin axisymmetric disk can be derived according to \cite[][page 73, Eq.~2-14b; Eq.~2-146]{Binney1987} and further \cite[][]{Conway2000, Gradshteyn1980}
\begin{equation} \label{e.disk_pot}
    \Psi_\mathrm{d}(r) = - 2 G \int_{r_\mathrm{in}}^{r_\mathrm{out}} \Sigma(r') \underbrace{k \, K(k) \sqrt{\frac{r'}{r}}}_{\raisebox{.5pt}{\textcircled{\raisebox{-.9pt} {1}}}} \, dr'
\end{equation}
and 
\begin{multline} \label{e.disk_pot_acc}
    \pop{\Psi_\mathrm{d}}{r} = \frac{G}{r} \int_{r_\mathrm{in}}^{r_\mathrm{out}} \Sigma(r') \\
    \underbrace{\left[ K(k) - \frac{1}{4} \Bigg(\frac{k^2}{1-k^2}\Bigg) \Bigg( \frac{r'}{r} - \frac{r}{r'}  \Bigg) E(k) \right] \sqrt{\frac{r'}{r}} k}_{\raisebox{.5pt}{\textcircled{\raisebox{-.9pt} {2}}}} \, dr' \:,
\end{multline}
where $r_\mathrm{in}$ and $r_\mathrm{out}$ represent the inner and outer boundary of the disk and $K(k)$ and $E(k)$ is the complete elliptical integral of the first and second kind respectively. The variable $k$ is defined as $k = \sqrt{4 r r' / (r + r')^2}$. For better readability we further substitute the terms \raisebox{.5pt}{\textcircled{\raisebox{-.9pt} {1}}} and \raisebox{.5pt}{\textcircled{\raisebox{-.9pt} {2}}} with $M_\mathrm{ij}$ and $N_\mathrm{ij}$ respectively. Since $\Sigma$ is constant within each cell per timestep, the discrete form of Eq.~\ref{e.disk_pot} and \ref{e.disk_pot_acc} can be written as
\begin{equation} \label{e.disk_pot_disc}
    \Psi_\mathrm{d}(r_{i}) =  \sum_{j=0}^{n_p-1}  \Sigma(r_{j}) \int_{r_{j}}^{r_{j+1}} M_{ij} dr_j
\end{equation}
and 
\begin{equation} \label{e.disk_pot_acc_disc}
    \pop{\Psi_\mathrm{d}(r_i)}{r} =  \sum_{j=0}^{n_p-1}  \Sigma(r_{j}) \int_{r_{j}}^{r_{j+1}} N_{ij}  dr_j \:.
\end{equation}
The indices $i$ and $j$ represent the $i$-th and $j$-th grid points and $n_p$ the total number of grid points within the computational domain. The individual integrals $\int_{r_{j}}^{r_{j+1}}$ are numerically solved via a trapezoidal rule. The remaining elliptical integrals $E(k)$ and $K(k)$ are approximated with the arithmetic-geometric mean (AGM) \cite[][]{Gerasimov1988}. A validation for the implemetation of $\Psi_\mathrm{d}$ and a more detailed discussion is presented in Sect.~\ref{ss.disk_potential_variation}.

\section{Numerical method}
\label{s.num}
\begin{figure}
    \centering
        \includegraphics[width=1.0\textwidth]{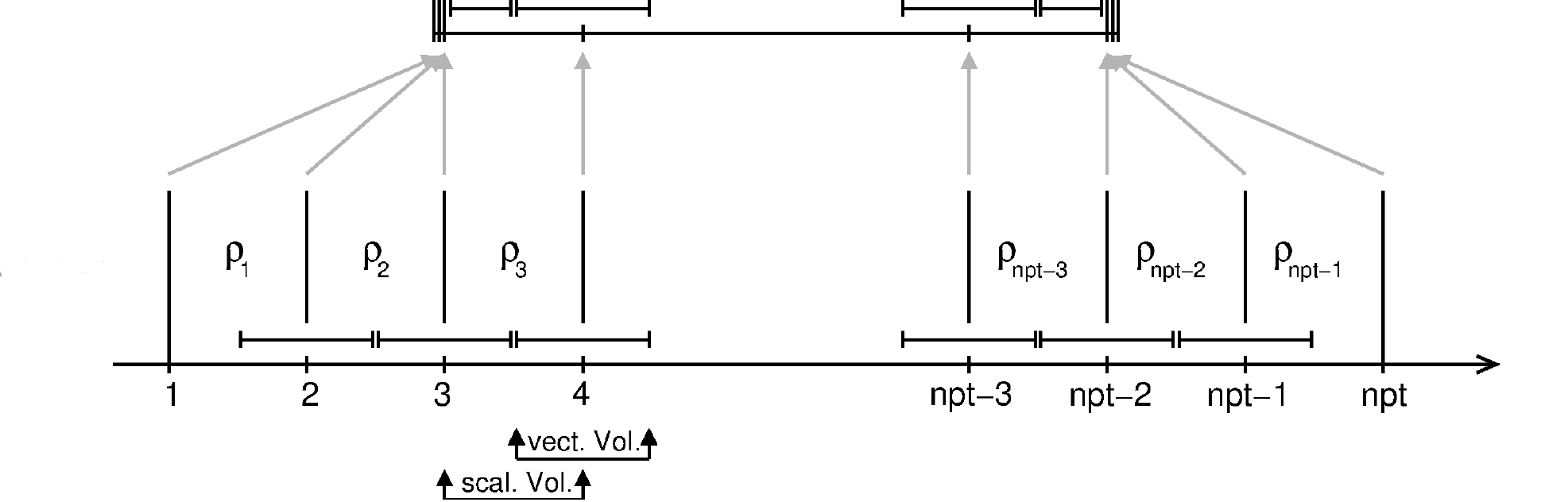}
    \caption{The overall picture of the discretization of an axial-symmetric configuration. The upper part of the figure shows the projection of this logical discretization scheme into the physical domain. The boundaries of the physical domain are located at $r_3$ and $r_\mathrm{npt-2}$, respectively. The cells between the grid points $r_1$ and $r_2$, or $r_\mathrm{npt-2}$ and $r_\mathrm{npt}$ have no physical volume. These cells are necessary to provide the boundary conditions for the model variables and are therefore referred to as `ghost cells'.}
    \label{f.tapir_grid}
\end{figure}

Since the method is based on the same principles as those presented in \cite{Dorfi1998} we restrict the discussion to specific issues typical for cylindrical geometry. The advection scheme is identical and based on a second-order advection method according to \cite{vanLeer1974}. Since we adopt a staggered mesh, we have to distinguish scalar and vector quantities. The scalar quantities are defined inside a finite volume and vector quantities are located at the cell boundaries. The exact locations are depicted in Fig.~\ref{f.tapir_grid}. Considering the shape of computational cells in cylindrical geometry, we have to take care of the corresponding divergence terms, use a volume-weighted density, and use the appropriate definition of the grid velocity through adaptive Reynolds theorem. Hence, the scalar volume $V_{\rm S}$ is defined by
\begin{equation}  \label{e.vol_s}
   V_\mathrm{S,i}(t)= \pi \left[  \rip^2(t) - r_i^2(t) \right] \:.
\end{equation}
For brevity, we omit from our notation the temporal dependence of the radial grid points $r_i$ in the remaining test. The flux term across a cell boundary is composed of two contributions resulting from the fluid flow with velocity $u_i$ and from a motion of the boundary itself, controlled by the adaptive grid, i.e.
\begin{equation} \label{e.Dvol_s}
    F_{\rm S} = 2\pi \, r_i \, u_i \, \delta t - \pi \delta\left( {r_i}^2 \right) \:,
\end{equation}
where the symbol $\delta$ denotes the temporal difference between of quantities between two timesteps.

For vector-like variables, the position within a cell is given by
\begin{equation}
   \rih^2 = \frac{1}{2} \left(  \rip^2 + r_i^2 \right)  \:,
\end{equation}
which demands a definition of the corresponding volume $V_{\rm V}$ for vector quantities through 
\begin{equation}  \label{e.vol_v}
  V_\mathrm{V,i} = \pi \left(  \rih^2 - r_{i-\frac{1}{2}}^2 \right) 
                      = \frac{\pi}{2} \left( \rip^2 - \rim^2 \right)  
\end{equation}
to ensure mass conservation.

In order to allow a correct treatment of the divergence terms within the adopted staggered cylindrical geometry, we calculate and conserve the flux by
\begin{equation}
   u_\mathrm{i+\frac{1}{2}}\rih = \frac{1}{2} \left(u_i r_i + u_{i+1}\rip\right)  \:,
\end{equation}
when transforming for the cell boundaries to the cell center and obtain the advection over the upper cell boundary at $r_\mathrm{i+\frac{1}{2}}$
\begin{equation} \label{e.Dvol_v}
    F_{\rm v} = \frac{2\pi}{2} \left( r_{i} \, u_{i} + r_{i+1} \, u_{i+1} \right)  \, \delta t
                 -          \frac{\pi}{2}  \delta \left( {r_{i}}^2 + {r_{i+1}}^2\right) \; .
\end{equation}

For a coordinate system moving with the velocity ${\bf u}^\mathrm{grid}$, we introduce the adaptive grid transport theorem of \cite{Winkler1984} which is a generalization of the well-known Reynolds transport theorem in hydrodynamics, given by
\begin{equation} \label{e.trafo}
   \frac{d}{dt}\left( \int_V f\,dV \right) 
   = \int_V \left[ \frac{\partial f}{\partial t} 
                  + \nabla\!\cdot\!\left(f{\bf u}^\mathrm{grid} \right)\right] dV .
\end{equation}
The motion of the grid points has to be defined in such a way that a constant function $f$ remains constant if the grid moves. Assuming only radial motions, we end up with an appropriate defintion of the grid velocity through
\begin{equation} \label{e.ugrid}
   2\pi  r {u_r}^\mathrm{grid}\delta t = \pi \delta\left( r^2 \right)  \:.
\end{equation}
Note that a Lagrangeian motion, i.e.~$u^{\mathrm{grid}} = u$, is then consistent with no fluxes across the cell boundaries.

\subsection{Discrete equations} \label{ss.discrete_equations}
The discretization of the hydrodynamic equations has been discussed in detail by \cite{Dorfi1998} and thus we refrain from defining the advected density $\tilde{\Sigma}$ which is obtained using a second order advection method \cite[see][]{vanLeer1974}. Accordingly, the derivation of the relative velocity $u^\mathrm{rel}$ between the grid cells and the gas is not part of this work. Furthermore, in order to present the hydrodynamic eqations in a more compact form, we are applying an index free notation and we denote the temporal differences with $\delta$ and the spatial differences with $\Delta$.

Applying Eq.~\ref{e.trafo} to Eq.~\ref{e.cont} we obtain
\begin{equation}
    \pop{}{t} \int_{V(t)} \Sigma\,dV + \int_{\partial V(t)} \Sigma u_r dS = 0 \:.
\end{equation}
Utilizing Eq.~\ref{e.vol_s} the discrete form of the equation of contiuity is given by
\begin{equation} \label{e.cont_disc}
    \delta[ \rho \; V_\mathrm{s}] + \sum \widetilde{\Sigma} \, \Delta\!{\rm Vol}_{\rm s} = 0  \:,
\end{equation}
where $\widetilde{\Sigma}$ denotes the surface density to be advected with the advective flux during the time step $\delta t$ over the cell boundary and therefore $\sum \widetilde{\Sigma} \, \Delta\!{\rm Vol}_{\rm s}$ represents the advection term.

Integrating Eq.~\ref{e.mot_r} over the volume and since ${\bf u}^\mathrm{grid}$ only has a radial component, the discretisation of the radial equation of motion takes the form
\begin{multline} \label{e.mot_r_disc}
    \delta[\overline{\Sigma} u_r \; V_\mathrm{v}]
        + \sum \widetilde{\overline{\Sigma} u_r} \, \Delta\!{\rm Vol}_{\rm v} 
        - \frac{\overline{\Sigma} \, u_\varphi u_\varphi}{r} \, V_{\rm v} \, \delta t
        + 2\pi r \sqrt{2\pi} \Delta \left[ H_\mathrm{p} P_\mathrm{gas} \right] \, \delta t \\
        - \partial_r \Psi(r)_{tot} \, \Sigma \, \Delta\!{\rm Vol}_{\rm v} \, \delta t
        - \frac{\pi}{r} \Delta \biggl\{ \overline{r}^2 \, \mu_\mathrm{Qr} \, \biggl[ \frac{\Delta u_r}{\Delta \overline{r}} - \frac{\overline{u}_r}{r} \biggr] \biggr\} \, \delta t = 0 \:,
\end{multline}
where $\partial_r \Psi(r)_{tot}$ tags the total radial gravitational acceleration (see Sect.~\ref{ss.disk_potential_variation}) and $\mu_\mathrm{Qr}$ is the radial viscosity coefficient. Quantities denoted with $\overline{F}$ are the averaged values transformed to the corresponding grid point in the staggered grid. The terms from left to right can be identified as the temporal difference, advection, the centrifugal force, the gas pressure gradient, gravitational acceleration and viscosity. To simplify the notation, we have abbreviated the radial difference between $r_i$ and $r_{i+1}$ by the spatial operator $\Delta$.

As a consequence of our 1D-method the velocity component $u_{\rm \varphi}$ is projected on the purely radial grid. Hence, Eq.~\ref{e.mot_phi} is discretized as a scalar-like quantity
\begin{equation}
    \delta[\overline{r} \Sigma u_\varphi \; V_{\rm s}]
        + \sum \widetilde{\overline{r} \Sigma u_\varphi} \, \Delta\!{\rm Vol}_{\rm s}
        - \pi \Delta \biggl\{ r^2 \, \mu_\mathrm{Q\varphi} \, \biggl[ \frac{\Delta u_\varphi}{\Delta \overline{r}} - \frac{\overline{u}_\varphi}{r} \biggr] \biggr\} \, \delta t = 0 \:,
\end{equation}
where $\mu_\mathrm{Q\varphi}$ is the angular scaling factor of the viscosity. In principle, $\mu_\mathrm{Qr}$ and $\mu_\mathrm{Q\varphi}$ can have different values, but presenting and validating the numerical method, we set them to identical values allowing a direct comparison with analytical viscous disk solutions (see Sect.~\ref{ss.viscous_evol}).

Equation~\ref{e.ene_ax} requires additional considerations. Since the internal energy balance of the disk is dominated by the irradiation term and the radiative cooling term, both implemented directly into the equation of internal energy, the temperature structure does not follow the radiation temperature (J), which in itself is computed correctly. The obvious solution for this problem is to implement the radiative energy flux also directly into the internal energy equation versus into the radiation energy equation, i.e.\ assuming $J=S$ (since the disk is optacally thick in radial direction). Further assuming an almost radial isotropic radiation field, i.e~ $f_\mathrm{edd} = 1/3$, we can drop the $3K-J$ term 
 and inserting Eq.~\ref{e.surf} leads to
\begin{equation}
    \sqrt{2\pi} \, H_\mathrm{p} \; 2\pi \Delta (r H_0) \, \delta t
        + \kappa_{\rm P} \, \Sigma \, (J_0-S) \, V_{\rm s} \, \delta t = 0 \:.
\end{equation}
Moreover, the time-dependent terms in the radiation flux equation are usually not important compared to the pure radiation diffusion terms. Hence, it makes sense to skip the time-dependent terms and the $3 K_{rr} - J$ and $H \, \nabla \cdot \bf{u}$ terms, which gives
\begin{equation}
    \frac{2}{3} r \, \sigma \, \Delta (T^4)
        + \overline{\kappa_{\rm R} \rho_0}  \, H_0 \, V_{\rm v} = 0 \:,
\end{equation}
leading to
\begin{equation}
    H_0 = - \frac{ \frac{2}{3} r \, \sigma \, \Delta (T^4) } { \overline{\kappa_{\rm R} \Sigma_0}  \, V_{\rm v} } \:.
\end{equation}
Utilizing the last equation, the discrete form of the equation of internal energy simplifies to
\begin{multline} \label{e.ene_disc}
    \delta[\Sigma e \, V_{\rm s}]
        + \sum \widetilde{(\Sigma  e)} \; \Delta\!{\rm Vol}_{\rm s}
        + P_\updownarrow \, 2\pi \Delta (r u_r) \, \delta t \\
        + 8\pi^2\sqrt{2\pi} \, H_p \, \sigma \; \Delta \left( - \frac{ \frac{2}{3} r^2 \, \Delta (T^4) }
                 { \overline{\kappa_{\rm R} \rho_0} \, V_{\rm v} } \right) \, \delta t \\
        + \Delta E\mathrm{rad} \\
        -\frac{\mu_{Q}}{2} \biggl\{  \biggl[ \frac{\Delta u_r}{\Delta r} - \frac{\overline{u}_r}{\overline{r}} \biggr]^2
        + \biggl[ \frac{\Delta u_\varphi}{\Delta r} - \frac{\overline{u}_\varphi}{\overline{r}} \biggr]^2 \biggr\}  V_{\rm s} \, \delta t = 0 \:,
\end{multline}
where the last term represents viscous energy generation due to friction. The updown arrow $\updownarrow$ denotes z-integrated values. The $\Delta E\mathrm{rad}$ stands for the radiation heating/cooling term in vertical direction $\Delta E_\mathrm{rad}$ (see Eq.~\ref{e.ene_ax}), which is discussed in detail in Sect.~\ref{s.heating_cooling}.

\subsection{Heating and cooling via stellar radiation}
\label{s.heating_cooling}
Radiation transport is a crucial component when simulating viscous disks because radiation heating and cooling leads to changes in the thermal disk profile and consequently to changes in the mass flux through the disk. Since astrophysical disks are predominantly optically thick in the radial direction, the radial radiation transport is only defined locally and thus has a minor impact on the long term evolution of disks. However, the radiative energy from the central star (and consequently the vertical radiation transport), becomes the dominant heating/cooling process. Our method adopts an axisymmetric geometry with vertically integrated values and therefore the vertical radiation transport needs a deeper discussion. 
The star's radiation hits the disk at a shallow angle that depends on the inclination of the disk and the stellar radius. The vertical location at which the star's radiation penetrates the disk is defined as $H_{\rm phot}$ and represents the vertical distance from the disk midplane to the photosphere of the gas disk. Therefore, the vertical scale height of the disk $H_{\rm p}$ and $H_{\rm phot}$ are constrained via a constant value $\alpha_{\rm phot}$ 
\begin{equation} \label{e.H_phot}
    H_{\rm phot} = \alpha_{\rm phot}\,H_{\rm p} \:.
\end{equation}
The area which is exposed to solar radiation is given by $2\pi r^2 \Delta(H_{\rm phot}/r)$, where $\Delta(H_{\rm phot}/r)$ represents the inclination of the surface area to the line-of-sight of the central star. Thus, the irradiation heating rate from the star is
\begin{equation}
    \partial_t E_{\rm irr} = 2(1-a)\frac{L_\star}{4\pi r^2}\, 2\pi r^2 \Delta\left( \frac{H_{\rm phot}}{r} \right) \:,
\end{equation}
where $a$ repersents the disk's albedo and $L_\star$ the stellar luminosity. The factor $2$ appears when taking into account that the disk is radiated from both the top and bottom. Utilizing the relation in Eq.~\ref{e.H_phot}, combining all constants in a factor $f_{\rm irr} = (1-a)\alpha_{\rm phot}$ and, in order to prevent negative values for $E_{\rm irr}$, the lower limit of $E_{\rm irr}$ per timestep $\delta t$ gives
\begin{equation} \label{e.E_irr_no_star}
    E_{\rm irr} = L_\star f_{\rm irr}\,max\left[ \Delta \left( \frac{H_{\rm p}}{r} \right),0 \right] \:.
\end{equation}
Furthermore, the physical size of the central star has to be taken into account as the stellar radiation is assumed to originate from a certain distance $H_\star$ from the midplane of the disk. The distance $H_\star$ can be computed from a half circle with stellar radius $R_\star$ using
\begin{equation} \label{e.H_star}
    H_\star = \frac{4}{3\pi}\,R_\star \:.
\end{equation}
Implementing Eq.~\ref{e.H_star} into Eq.~\ref{e.E_irr_no_star} leads to the radiation enrgy from a star on the disk surface at a certain orbital radius, given by
\begin{equation}
    E_{\rm irr} = L_\star f_{\rm irr}\,max\left[ \Delta \left( \frac{H_{\rm p} - H_\star}{r} \right),0 \right] \:.
\end{equation}

To describe the radiative cooling of the disk, we assume that the radiation of a gid-cell volume $V_{\rm s}$ can be specified by the black body radiation and hence the cooling energy $E_{\rm cool}$ per timestep is
\begin{equation}
    E_{\rm cool} = 2 V_{\rm s} \sigma T_{\rm surf}^4 \delta t \:,
\end{equation}
where $\sigma$ is the Stefan-Boltzmann constant and $T_{\rm surf}$ the local surface temperature of the disk. Taking into account that the heating energy from the star and the cooling energy from the disk have to be transported in the $z$-direction, the radiative energy balance $\Delta E_{\rm rad}$ (see. Eq.~\ref{e.ene_ax}) is
\begin{equation}
    \Delta E_{\rm rad} = 8\pi H_zV_{\rm s} \delta t \:.
\end{equation}
The final step is to calculate the radiation flux in the vertical direction $H_z$ which is illustrated in Sect.~\ref{ss.radiation_flux_in_z}.

\subsubsection{Radiation flux in z-direction}
\label{ss.radiation_flux_in_z}
In order to compute the radiation flux in the vertical direction, we utilize the stationary radiation transport equation
\begin{equation}
    {\bf n} \cdot {\bf \nabla} I_{\nu} = \eta_{\nu} - \chi_{\nu}I_{\nu} \:.
\end{equation}
Integration over all directions and frequencies gives the radiation energy equation (zero$^{\rm th}$ moment)
\begin{eqnarray}
    \frac{1}{4\pi} \int d\Omega: & \qquad & {\bf \nabla} \cdot {\bf H_{\nu}} = \eta_{\nu} - \chi_{\nu}J_{\nu} \nonumber \\
    \int d\nu:                   & \qquad & {\bf \nabla} \cdot {\bf H = \eta} - \chi_{\rm P} J \stackrel{\eta = \chi S}{=} \chi_{\rm P} (S-J) \nonumber \\
                                 & \qquad & \pop{H_{z}}{z} = - \kappa_{\rm P} \rho \, (J-S)
\end{eqnarray}
while integration with $\bf n$ yields the first moment equation of radiation
\begin{eqnarray}
    \frac{1}{4\pi} \int {\bf n} d\Omega: & \qquad & {\bf \nabla} \cdot \mathbf{K}_{\nu} = - \chi_{\nu} {\bf n} I_{\nu} = - \chi_{\nu} {\bf H_{\nu}} \nonumber \\
    \int d\nu:                          & \qquad & {\bf \nabla} \cdot \mathbf{K} = - \chi_{\rm R} H_z \nonumber \\
                                        & \qquad & \pop{K_{zz}}{z} = - \chi_{\rm R} H_z = - \kappa_{\rm R} \rho \, H_z \:.
\end{eqnarray}
In proceeding further we list and apply our idealizing assumptions:
\begin{description}
    \item[Eddington factor:] $\mathbf{K} = \mathbf{f}_{\rm edd} J$, or in z-components $K_{zz} = f_{\rm edd} J = \dfrac{1}{3} J$.
    \item[Local thermal equilibrium (LTE):] $J = S = \dfrac{\sigma}{\pi} T^4$
\end{description}

Integration of the radiation energy equation thus gives
\begin{equation}
   \pop{H_{z}}{z} = - \kappa_{\rm P} \rho \, (J-S) = 0 \quad \longrightarrow \quad \textrm{H = const.}
\end{equation}

Integration of the radiation flux equation yields
\begin{eqnarray} \label{}
   \pop{K_{zz}}{z} &=& \dfrac{1}{3} \pop{J}{z} = - \kappa_{\rm R} \rho \, H_z \nonumber \\
                  &=& \dfrac{1}{3} \dfrac{\sigma}{\pi} \pop{T^4}{z} = - \kappa_{\rm R} \rho \, H_z \nonumber \\
                  &=& \dfrac{1}{3} \dfrac{\sigma}{\pi} \frac{T_{\rm surf}^4-T_0^4}{H_\mathrm{p}} = - \kappa_{\rm R} \rho_0 \, H_z
\end{eqnarray}

The position of the disk's photosphere $H_{\rm phot}$ would be the proper z-localization for the surface temperature $T_{\rm surf}$, however, the position of midplane temperature $T_0$ is less clear as the hydrodynamic discretization still assumes an isothermal structure. Therefore, using $H_{\rm p}$ as the spatial separation of $T_{\rm surf}$ and $T_0$ in the last equation above seems to be a reasonable compromise. Here we also specified the relevant density to be the midplane density $\rho_0$, since we only have one opacity which corresponds to the midplane, and the optical depth would likely be dominated by the dense layers close to the midplane.

Rewriting the last line, we obtain an expression for $H_{z}$, given by
\begin{equation}
   H_z = - \frac{\sigma}{3 \pi \kappa_{\rm R} \rho_0 H_{\rm p}} \left( T_{\rm surf}^4-T_0^4 \right) \:.
\end{equation}
Identifying the optical depth in z direction as $\tau = \kappa_{\rm R} \rho_0 H_{\rm p}$ (i.e.\ $\tau = \kappa_{\rm R} \frac{\rho_\updownarrow}{\sqrt{2\pi}}$ ),
we obtain
\begin{equation}
   H_z = - \frac{\sigma}{3 \pi \tau} \left( T_{\rm surf}^4-T_0^4 \right)
\end{equation}
The contribution to the energy balance of the scalar cell $i$ in the time interval $\delta t$, computed from the $H_z$ flux above, is given by
\begin{eqnarray}
   \Delta E = \dfrac{4\pi}{c} \Delta J &=& 8\pi H_z \; V_{\rm s} \delta t \nonumber \\
                                       &=& - 4\pi \frac{2 \sigma}{3 \pi \tau} \left( T_{\rm surf}^4-T_0^4 \right) \; V_{\rm s} \delta t \:,
\end{eqnarray}
where a factor $2$ enters to allow for radiation over both the upper and lower surfaces, and the factor $4\pi$ relates the
first moment of radiation $H$ to the energy flux $F_{\rm rad}$. In order to treat the flux direction correctly, we assume the radiation \textit{from the disk to the surrounding space} as positive for $H_z$, a positive $\Delta E$ and therefore the cell loses energy (i.e.\ $\delta[\rho e] + \Delta E + \ldots = 0$, or $\delta[J] + \frac{c}{4\pi} \Delta E + \ldots = 0$, respectively).

\subsubsection{Surface temperature of the disk}
\label{ss.surface_temp_of_disk}
To calculate the vertical energy flux and thus the contribution to the internal energy equation, we have to compute an estimate for the surface temperature, $T_{\rm surf}$, of the disk. In order to do so, we assume local balance between irradiation and radiative cooling on the disk surface using the description from Sect.~\ref{s.heating_cooling}. Additionally, we consider the vertical energy transport from the disk interior to the disk surface which corresponds to the irradiation/cooling term $\Delta E_{\rm rad}$ in the internal energy equation (Eq.~\ref{e.ene_ax}) and in the following, we denote this quantity $F_{\rm vert}$. Hence, the energy balance for the disk surface is
\begin{equation}
    F_{\rm vert} + E_{\rm irr} + E_{\rm amb} - E_{\rm cool} = 0 \:,
\end{equation}
where $E_{\rm amb}$ is the irradiation from the ambient radiation field for which we assume, as for the radiative cooling $E_{\rm cool}$, black body radiation. The contributions then is
\begin{eqnarray}
    F_{\rm vert} &=& \frac{8 \sigma}{3 \tau} \left( T_0^4 - T_{\rm surf}^4 \right) V_{\rm s} \, \delta t \:, \\
    E_{\rm irr}  &=& L_\star \, f_{\rm irr} \, \max \left[ \Delta{\left( \frac{H_p - H_\star}{r} \right)},0 \right] \delta t \:, \\
    E_{\rm amb}  &=& 2 V_{\rm s} \, \sigma T_{\rm amb}^4  \, \delta t \:, \\
    E_{\rm cool} &=& 2 V_{\rm s} \, \sigma T_{\rm surf}^4 \, \delta t \:.
\end{eqnarray}
Summing these terms up, skipping $\delta t$, and dividing by $2 \sigma V_{\rm s}$ gives
\begin{equation*}
   \frac{4}{3 \tau} \left( T_0^4 - T_{\rm surf}^4 \right)
    + \frac{L_\star \, f_{\rm irr}}{2 \sigma V_{\rm s}} \, \max \left[ \Delta{\left( \frac{H_{\rm p} - H_\star}{r} \right)},0 \right]
    + T_{\rm amb}^4
    - T_{\rm surf}^4 = 0
\end{equation*}
which can be rewritten as
\begin{equation*}
   \frac{4}{3 \tau} T_0^4
    + \frac{L_\star \, f_{\rm irr}}{2 \sigma V_{\rm s}} \, \max \left[ \Delta{\left( \frac{H_p - H_\star}{r} \right)},0 \right]
    + T_{\rm amb}^4
    = T_{\rm surf}^4 \left( 1 + \frac{4}{3 \tau} \right) \:.
\end{equation*}
This leads to an expression for $T_{\rm surf}^4$, given by
\begin{equation}
T_{\rm surf}^4 =
   \frac{\frac{4}{3 \tau}}{1 + \frac{4}{3 \tau}} T_0^4
    + \frac{1}{1 + \frac{4}{3 \tau}}  \frac{L_\star \, f_{\rm irr}}{2 \sigma V_{\rm s}} \, \max \left[ \Delta{\left( \frac{H_p - H_\star}{r} \right)},0 \right]
    + \frac{1}{1 + \frac{4}{3 \tau}}  T_{\rm amb}^4 \:,
\end{equation}
Since we have adopted the Eddington-approximation for calculating the vertical temperature structure, the optical thin limit usually overestimates of the radiative losses. In order to correct this feature a flux-limited diffusion approximation can be utilized \citep[e.g.][]{Mihalas1984}. The radiative energy losses can be computed for both, the optically thin and optically thick case, by rewriting the terms with $\tau$ as
\begin{eqnarray*}
    \frac{\frac{4}{3 \tau}}{1 + \frac{4}{3 \tau}} = \frac{1}{1 + \frac{3 \tau}{4}} \:, \\
    \frac{1}{1 + \frac{4}{3 \tau}}                = \frac{\frac{3 \tau}{4}}{1 + \frac{3 \tau}{4}} \:.
\end{eqnarray*}
Insertion of the result for $T_{\rm surf}^4$ into the equation for the vertical radiation flux gives
\begin{eqnarray}
   \Delta E_{\rm rad} = \frac{8 \sigma}{3 \tau} \left( T_0^4 - T_{\rm surf}^4\right) V_{\rm s} \, \delta t \:.
\end{eqnarray}
This leads to
\begin{multline*}
   \Delta E_{\rm rad} = \frac{8 \sigma}{3 \tau} \left( T_0^4
    - \frac{1}{1 + \frac{3 \tau}{4}} T_0^4 \right. \\
    - \frac{\frac{3 \tau}{4}}{1 + \frac{3 \tau}{4}}  \frac{L_\star \, f_{\rm irr}}{2 \sigma V_{\rm s}} \, \max \left[ \Delta{\left( \frac{H_{\rm p} - H_\star}{r} \right)},0 \right] - \frac{\frac{3 \tau}{4}}{1 + \frac{3 \tau}{4}}  T_{\rm amb}^4
                                      \biggr) V_{\rm s} \, \delta t \:.
\end{multline*}
Using
\begin{equation*}
    1 - \frac{1}{1 + \frac{3 \tau}{4}} = \frac{\frac{3 \tau}{4}}{1 + \frac{3 \tau}{4}} \:,
\end{equation*}
and rewriting the leading factor, gives
\begin{multline}
   \Delta E_{\rm rad} = 2 \sigma V_{\rm s} \frac{1}{ \frac{3 \tau}{4} } \left( \right.
   \frac{\frac{3 \tau}{4}}{1 + \frac{3 \tau}{4}}  T_0^4 \\
    - \frac{\frac{3 \tau}{4}}{1 + \frac{3 \tau}{4}}  \frac{L_\star \, f_{\rm irr}}{2 \sigma V_{\rm s}} \, \max \left[ \Delta{\left( \frac{H_p - H_\star}{r} \right)},0 \right]
    - \frac{\frac{3 \tau}{4}}{1 + \frac{3 \tau}{4}}  T_{\rm amb}^4
                                                     \bigg) \delta t \:.
\end{multline}
By canceling $\frac{3 \tau}{4}$ and introducing a modified optical depth $\tau'$ with the definition
\begin{equation}
    \tau' = \frac{3 \tau}{4} \left( = \frac{3}{4} \kappa_{\rm R} \rho_0 H_p \right)
\end{equation}
the result can be written in the form of
\begin{multline} \label{e.E_rad_opt_thick}
   \Delta E_{\rm rad} = 2 \sigma V_{\rm s} \frac{1}{1+\tau'} \left( T_0^4 - T_{\rm amb}^4 \right) \delta t \\
            - \frac{1}{1+\tau'} L_\star \, f_{\rm irr} \, \max \left[ \Delta{\left( \frac{H_p - H_\star}{r} \right)},0 \right] \delta t \:.
\end{multline}

Note that Eq.~\ref{e.E_rad_opt_thick} is valid if the disk is optically thick. As the optically thick part of the irradiation is always dominant, the optically thin part can be neglected (if we reach the optically thin limit, we simply have an isothermal structure). Nevertheless, we have to account for the weakening of the irradiation energy when approaching the optically thin limit. This can be achieved by adding a weighted factor to Eq.~\ref{e.E_rad_opt_thick}. For the optically thick limit, we can define
\begin{equation*}
   \frac{\tau^n}{1+\tau^n} \rightarrow \left\{ \begin{array}{cl}
                                 0 & \quad \mbox{for} \; \tau \ll 1 \:,  \\
                                 1 & \quad \mbox{for} \; \tau \gg 1 \:.
                    \end{array} \right.
\end{equation*}
In the optically thin case, we have $\tau \rightarrow 0$ and thus
\begin{eqnarray*}
1 + \tau^n &\rightarrow& 1 \:, \\
1 + \tau'  &\rightarrow& 1 \:.
\end{eqnarray*}
Therefore, to account for an optical thin disk, the exponent must satisfy $n \ge 2$. Utilizing $\tau' = 3/4 \tau$ leads to the final form of the heating/cooling input into the disk
\begin{multline} \label{e.E_rad_final}
    \Delta E_{\rm rad} = \frac{4}{3} \frac{\tau}{1+\tau^2} \delta t \biggl(
        \sigma V_{\rm s} \frac{1}{1+\tau'} \left( T_0^4 - T_{\rm amb}^4 \right)  \\
        - L_\star \, f_{\rm irr} \, \max \left[ \Delta{\left( \frac{H_p - H_\star}{r} \right)},0 \right] \biggr) \:.
\end{multline}

\subsection{Adaptive grid}
\label{s.grid}
To provide a sufficient radial resolution, we use an adaptive grid which is augmented to the physical equations and redistributes the gridpoints at every timestep. Following the strategy of \cite{Dorfi1987}, where all details of the adaptive grid are presented, the basic parameters are controlling the spatial variation and the temporal smoothing of the grid motion. In particular, the temporal grid-scale $\tau_\mathrm{g}(r)$ is typically set to the global diffusive timescale $\bar\tau_\nu$ (see Sect.~\ref{ss.viscous_evol}) and we use $\tau_\mathrm{g} = \bar\tau_\nu$. Since the spatial variation of the physical variables is expected to be rather smooth, only the surface density $\Sigma$ enters the desired resolution.

\begin{figure}[]
    \centering
    \includegraphics[width=0.9\textwidth]{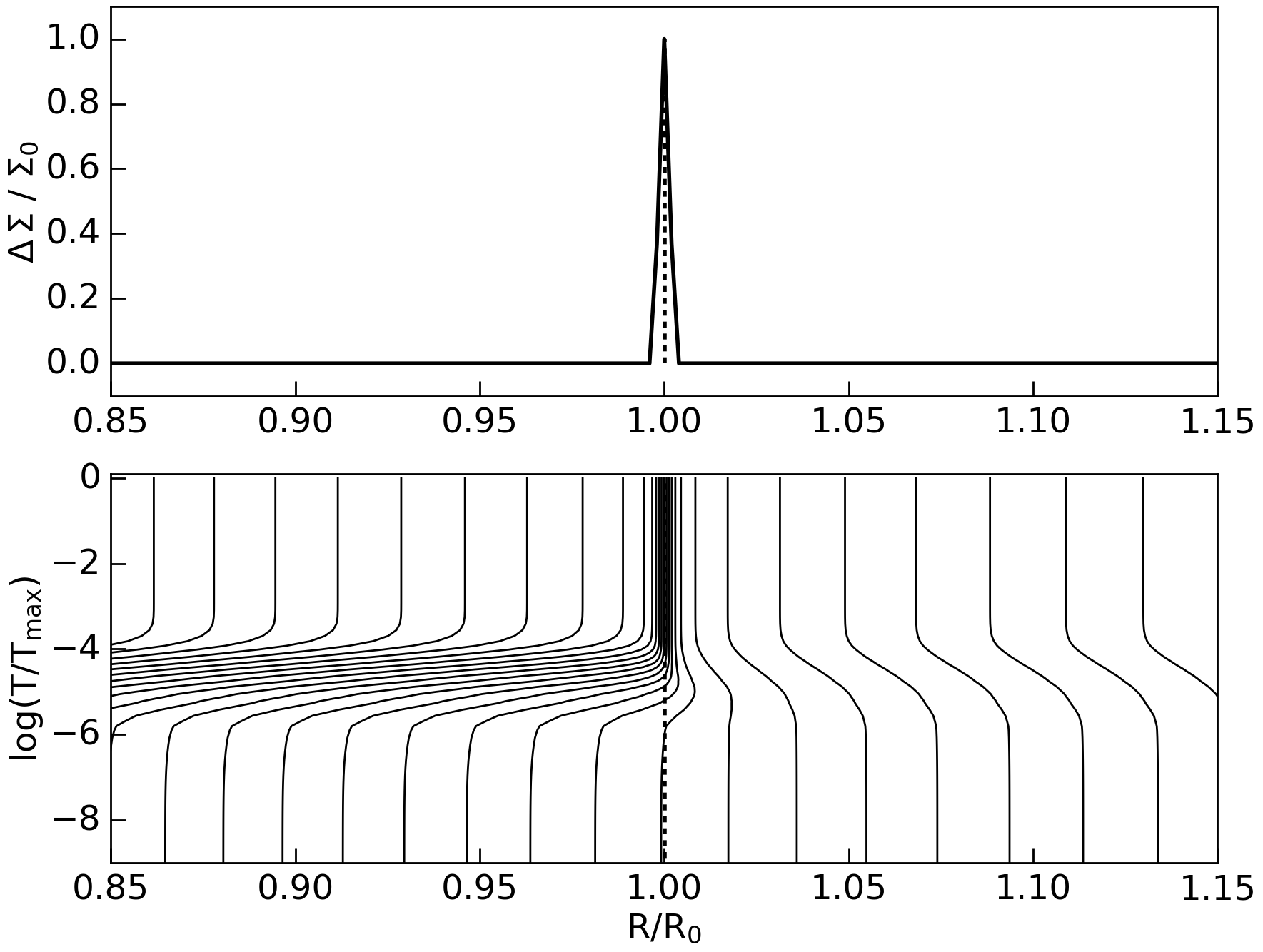}
    \caption{Propagation of gridpoints towards a density feature imposed on a homogeneous background. The upper panel shows the surface density $\Sigma$ in values of a reference density $\Sigma_0$ against the radius $R$ in values of a reference radius $R_0$. The density feature has a maximum of $\Sigma_0$ at the location of $R_0$. Starting from an initially logarithmic equidistant gridpoint distribution, the gridpoints accumulate around the density feature to increase the local resolution (lower panel).}
    \label{f.disk_grid}
\end{figure}

In Fig.~\ref{f.disk_grid} we illustrate the redistribution of gridpoints towards an artificial density feature. Starting from an initial model with a homogeneous density profile on which we impose a disturbance, $\Sigma_0$, at a certain radial distance $R_0$ (top panel). After initiating the simulation and only solving the grid equation, one can see that the initially logarithmic equidistant gridpoints move towards the density peak leading to a higher gridpoint concentration around the disturbance (bottom panel). This behaviour demonstrates the dynamic improvement of accuracy at regions with physical features, e.g. wave-fronts.

\subsection{Numerical solution procedure}
Summing up all radial indices of the discrete equations (see last sections), we find that the difference equations contain variables at $5$ gridpoints, i.e.~the discrete equations connect $[i-2, i-1,i,i+1,i+2]$. This 5-point stencil corresponds to a Penta-diagonal block-structured Jacobi-matrix, which has to be inverted during the Newton-Raphson iteration for solving the non-linear system of equations. Each block-submatrix is a quadratic $m\times m$-matrix, where $m$ is the number of unknowns per gridpoint. In the simplest case, we have $m=7$ with the variables being $\Sigma, u_r, \uf, e, H, \Psi_{\mathrm d}$ and $r$ and for 500 radial gridpoints the whole system of algebraic equations contains $7\times 500=3500$ unknowns to be calculated and updated at each timestep. Usually, these iterations converge after a few cycles to a relative accuracy $\delta_x$ of $\delta_x > 10^{-6}$. More details on these numerical solution issues can be found in \cite{Dorfi1998}. Boundary conditions according to Sect.\ref{ss.BC} are implemented through ghost cells to keep the overall matrix pattern simple.

\section{Numerical examples, accuracy and comparison to analytical solutions}
\label{s.example}
To assess the validity of our model when it is applied to astrophysical viscous disks and to meet their specific requirements mentioned in the previous sections, we tested our code against several theoretical models. In Sect.~\ref{s.intro}, we emphasized the importance of the inner boundary condition and hence in Sect.~\ref{ss.BC}, we show that leaving out the very inner regions of the disk leads to an entirely different disk structure. Thus results from models with large disk-to-star gaps have to be treated carefully. In general, defining an exact outer boundary for the disk is difficult as there is a pressure gradient towards the surrounding ISM. Our model has no restrictions of where to set the outer boundary and hence we can cover several orders of magnitude of disk radius.
As viscous disks are defined via a thermal and gravitational profile, permuting the disk in either the thermal or the gravitational variables should lead to a wave propagation in either of the variable spaces (temperature-radius and density-radius space). In Sect.~\ref{ss.thermal_adjustment} and \ref{ss.density_waves}, we show that our model reproduces the expected behaviour which validates the model setup.
Viscosity plays an essential role in astrophysical disks as it determines the mass accretion onto the star. Since we have implemented a viscosity model (see Sect.~\ref{s.phys_equations}), we test our model against an analytical solution (cf. Subsect.~\ref{ss.viscous_evol}).

\subsection{Boundary conditions}
\label{ss.BC}
As mentioned in Sect.~\ref{s.intro}, we emphasize that the boundary conditions play an important role in the long term evolution of astrophysical, viscous disks and hence we test our model accordingly. We have stated that changing the inner boundary of the disk consequently leads to entirely different disk structures whereas varying the outer boundary, the disk structure remains similar. To verify this, we start from a reference disk with a fixed inner and outer radius ($R_\mathrm{in}=0.05$AU, $R_\mathrm{out}=25$AU) and vary 1) the inner radius and leave the outer disk radius constant and 2) the outer radius and leave the inner radius constant. The radii variations and the dimensions of the reference disk are presented in Tab.~\ref{t.boundary_conditions}. For this test, we have modified the boundary conditions by $1/2 R_\mathrm{ref}$ (min values) and $10 \times R_\mathrm{ref}$ (maximum values). Note that the exact radii values are not important for the qualitative simulation result but should be mentioned for completeness.

\begin{table}[]
    \centering
    \begin{tabular}{lcc}
    \textbf{}                     & \textbf{inner radius} & \textbf{outer radius} \\
    reference disk                & 0.05 AU               & 25 AU                 \\
    $R_\mathrm{in;min}$           & 0.025 AU              & 25 AU                 \\
    $R_\mathrm{in;max}$           & 0.5 AU                & 25 AU                 \\
    $R_\mathrm{out;min}$          & 0.05 AU               & 12.5 AU               \\
    $R_\mathrm{out;max}$          & 0.05 AU               & 250 AU                \\
    \end{tabular}
    \caption{List of boundary conditions tested with our model.}
    \label{t.boundary_conditions}
\end{table}

\begin{figure}[]
        \centering\includegraphics[width=\textwidth]{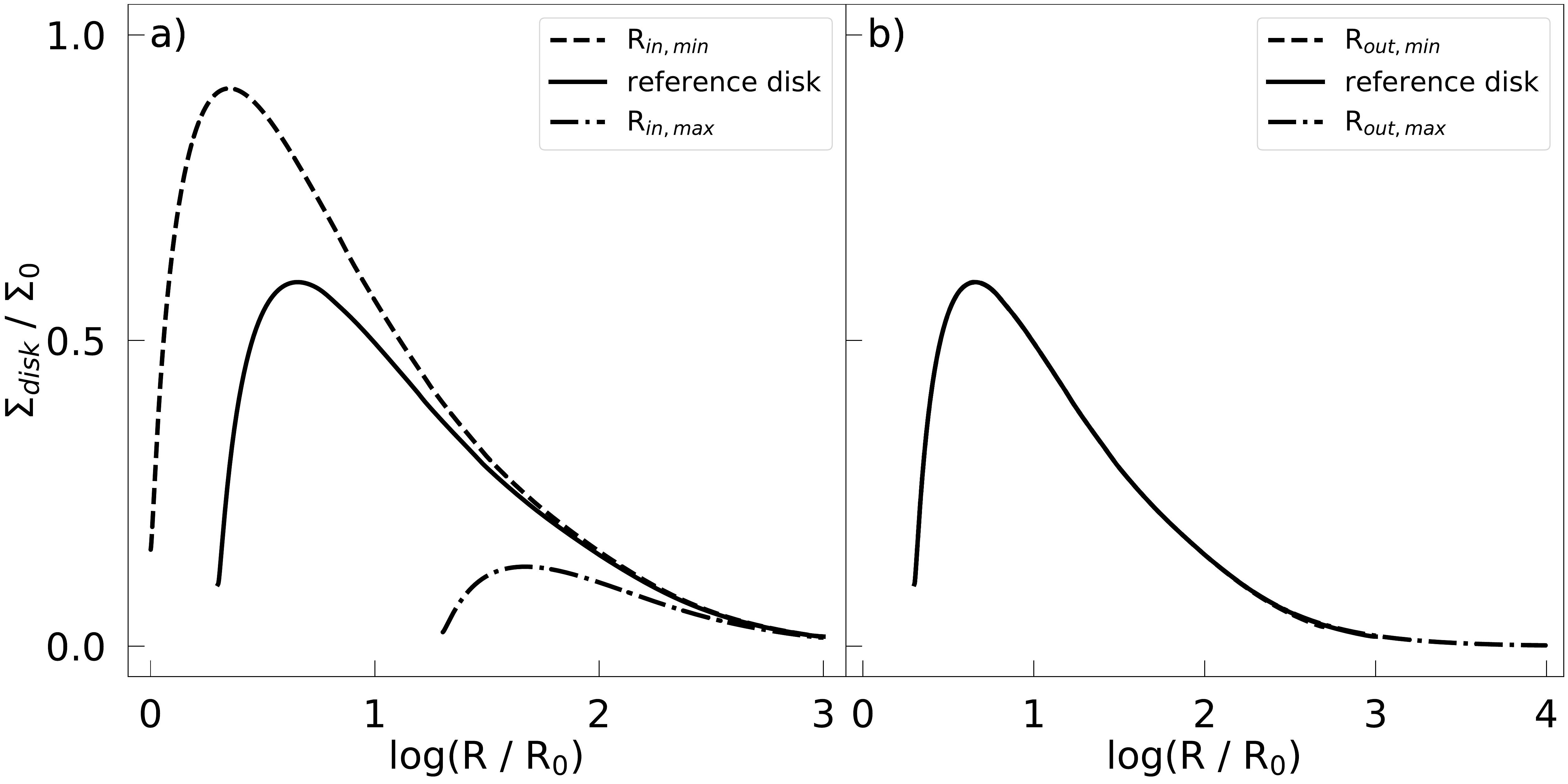}
        \caption{Surface density $\Sigma_\mathrm{disk}$ in values of an arbitrary reference surface density $\Sigma_0$ against disk radius $R$ in values of a reference radius $R_0$. Panel~a) shows the structure of a stationary viscous disk when only varying the inner boundary and panel~b) shows the structure of a disk when only varying the outer boundary and leave the inner boundary constant. In both cases, we modify the inner/outer radius by $1/2 R_\mathrm{ref}$ (dashed) and $10 R_\mathrm{ref}$ (dashed-dot), where $R_\mathrm{ref}$ represents the inner/outer radius of the reference disk. In both cases, all models have equal mass flux. One can see that varying the inner boundary leads to entirely different disk structures whereas when changing the outer disk radius, the disk remains unaltered.}
        \label{fig.IBCtest}
\end{figure}

To make disks comparable, the mass flux of the different models must be equal as the flux is entirely determined by viscosity. In Fig.~\ref{fig.IBCtest}, we plot the disk's surface density profile against the radius in dimensionless variables. Panel~a) shows simulations in which the inner boundary is varied with a constant outer boundary and panel~b) shows simulations in which the outer boundary is varied with a constant inner boundary. We assume pressureless outflow, $\partial P / \partial r = 0$, at the inner boundary which results in the drop of the density profile towards the centre of the disk. In reality, radiation pressure and magnetic pressure would be present but as this physical boundary has not yet been investigated, we assume $P=0$ as the lower boundary. The main argument for this assumption is that we have a real starting point of a disk and we do not need to assume a smart-cell \cite{Vorobyov2018, Vorobyov2019} which can not account for physical processes in the innermost disk regions. From Fig.~\ref{fig.IBCtest}, Panel~a), we see that modifying the inner radii from $1/2$ to $10$ times the inner radius of the reference disk (solid line, compare Tab.~\ref{t.boundary_conditions}) leads to an entirely different surface density profile. As the surface density is tightly coupled to the viscosity, such a variation must result in a different evolution track for the disk (e.g FU Ori bursts, Ex Lup objects, etc.).

If varying the outer boundary condition, represented by Panel~b), we see that the disk structure stays unaffected. This outcome also supports the assumption that the density in the outer regions of the disk is of the order of the ISM density ($\rho_\mathrm{ISM} = 10^{-24}$ gcm$^{-3}$) \cite{McKee1977} and therefore viscous drag is negligible. We emphasize that our model can cover large spatial dimensions, meaning that we do not have any limits in where to set the outer boundary. For numerical accuracy and simulation time optimisation, we set the outer boundary to that of the outer base radius according to the standard disk model \cite{Armitage2013}.

\subsection{Thermal adjustment}
\label{ss.thermal_adjustment}
When increasing the host stars luminosity, the thermal pressure in the disk will increase due to radiation. The higher pressure breaks hydrostatic equilibrium in the vertical disk structure and thus the disk will expand in the vertical direction. Changes in the disk dynamics can only appear on the viscous timescale of the disk which increases with $\tau_\mathrm{\nu} = r^2/\nu$ further from the star. Additionally, the surface temperature of the disk decreases with increasing distance from the centre and as a result, the innermost disk regions are expected to adjust faster to the star's thermal input compared to regions further away. On the contrary, due to the inwards-oriented accretion flow, regions closer to the star are cooled by cold gas from the outer regions of the disk. Hence, the local $\tau_{\nu}$ decreases when getting closer to the centre which leads to a delay in the thermal adjustment of these disk regions. Outer disk regions are less affected and therefore adjust faster to the thermal input of the star and as a consequence, the thermal profile of the disk is expected to approach the final state from both inwards and outwards, with less effective zones in the middle of the disk.

\begin{figure}[]
        \centering\includegraphics[width=\textwidth]{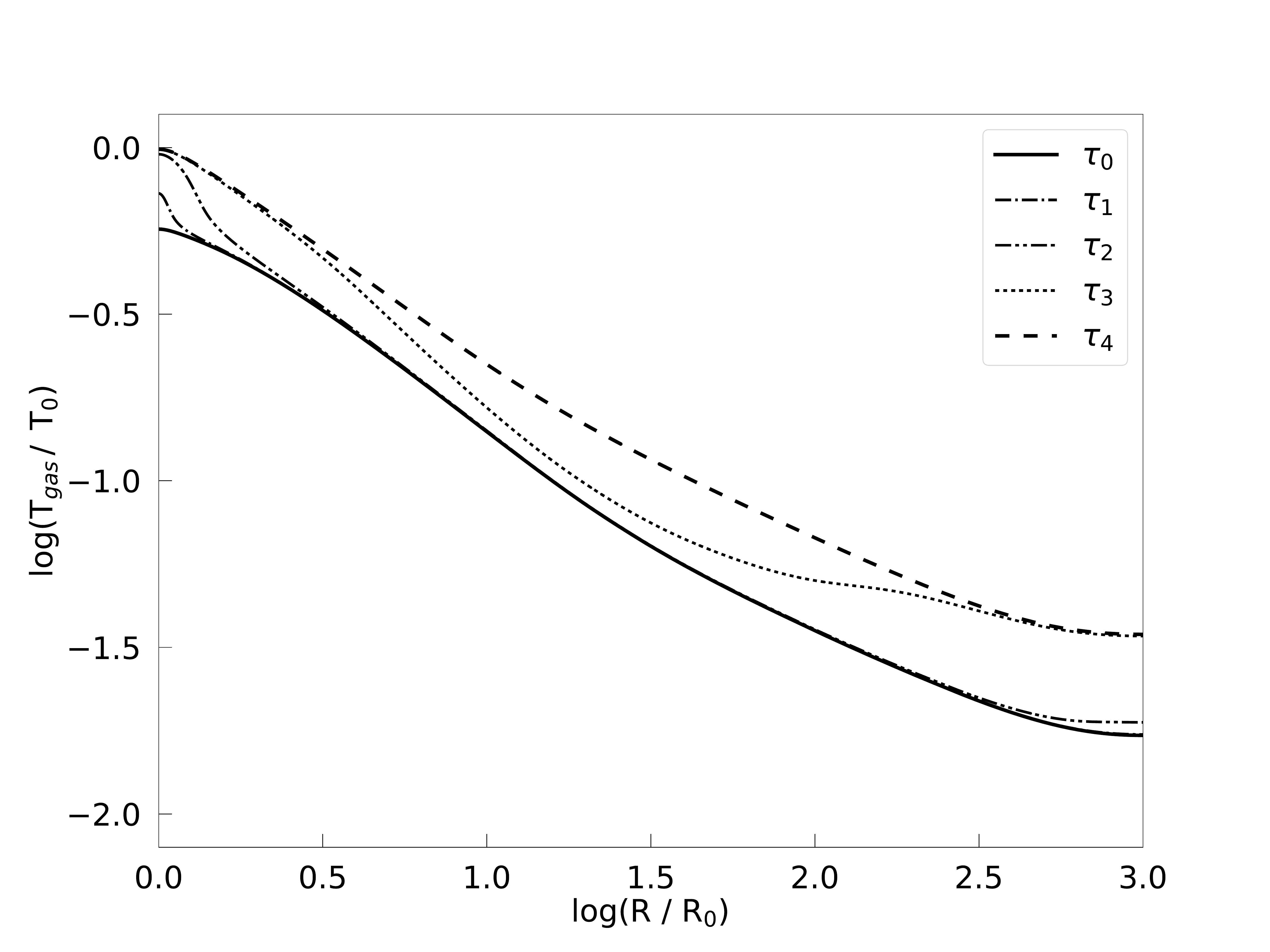}
        \caption{Temperature evolution of a viscous disk when spontaneously increasing the host star's luminosity. We show the gas temperature against the radius in dimensionless values. The initial model is illustrated by $\tau_0$ (solid) and the final model after changing the luminosity of the central star is shown by $\tau_4$ (dashed). One can see that the innermost grid cells respond quickly to the updated thermal input ($\tau_\mathrm{1}$, dashed-dotted). Since the viscous timescale $\tau_\nu$ decreases with distance from the centre and because of accretion flow from the outer cold regions of the disk, the temperature gradually adapts to $\tau_4$ from the outer and inner parts of the disk ($\tau_2$, $\tau_3$).}
        \label{f.thermal_waves_tgas}
\end{figure}

To test our model accordingly, we started from a stationary viscous disk around a Sun-like star. We then spontaneously increased the stars luminosity by an arbitrary factor $\Delta L$. We found that regardless of the exact value of $\Delta L$, the qualitative outcome of our simulation remains similar and therefore all our simulations have been carried out with $\Delta L = 10$ for illustration purposes. In Fig.~\ref{f.thermal_waves_tgas}, we show the thermal evolution of the disk from the initial model ($\tau_\mathrm{0}$, solid) to the final stationary solution ($\tau_\mathrm{4}$, dashed). The plot shows the gas temperature as a function of the disk radius. As expected, the inner region responds immediately to the updated thermal input from the host star ($\tau_\mathrm{1}$, dashed-dotted) as the very inner disk-cells contain less material according to the pressureless outflow condition at the inner boundary (compare Sect.~\ref{ss.BC}). Consequently, the local viscosity is less effective and additionally the thermal input of the star is near the maximum. Therefore, the disk gas temperature can adjust to the updated thermal input on a short timescale. According to the radially increasing $\tau_\nu$, the temperature only gradually increases with distance from the disk centre ($\tau_\mathrm{2}$, dash-dot dotted) and due to mass accretion from the outer still cold disk regions, the local temperature adjustment is slowed down significantly at around $30 \,R_0$. Simultaneously, the very outer grid cell is not affected by accretion of cold gas and thus the temperature can freely adjust to the radiation from the star ($\tau_\mathrm{3}$, dotted). The thermal profile of the disk then adapts gradually from the outer and the inner regions until the final stationary solution is reached. We emphasize that the timescales on which the thermal adjustments occur ($\tau_0$ to $\tau_4$) are about three times the order of the mean Keplerian rotation period and hence utilizing an implicit scheme is necessary in order to get reasonably large timesteps to investigate the full lifespan of astrophysical viscous disks.

\subsection{Density waves}
\label{ss.density_waves}
\begin{figure}[]
        \centering\includegraphics[width=\textwidth]{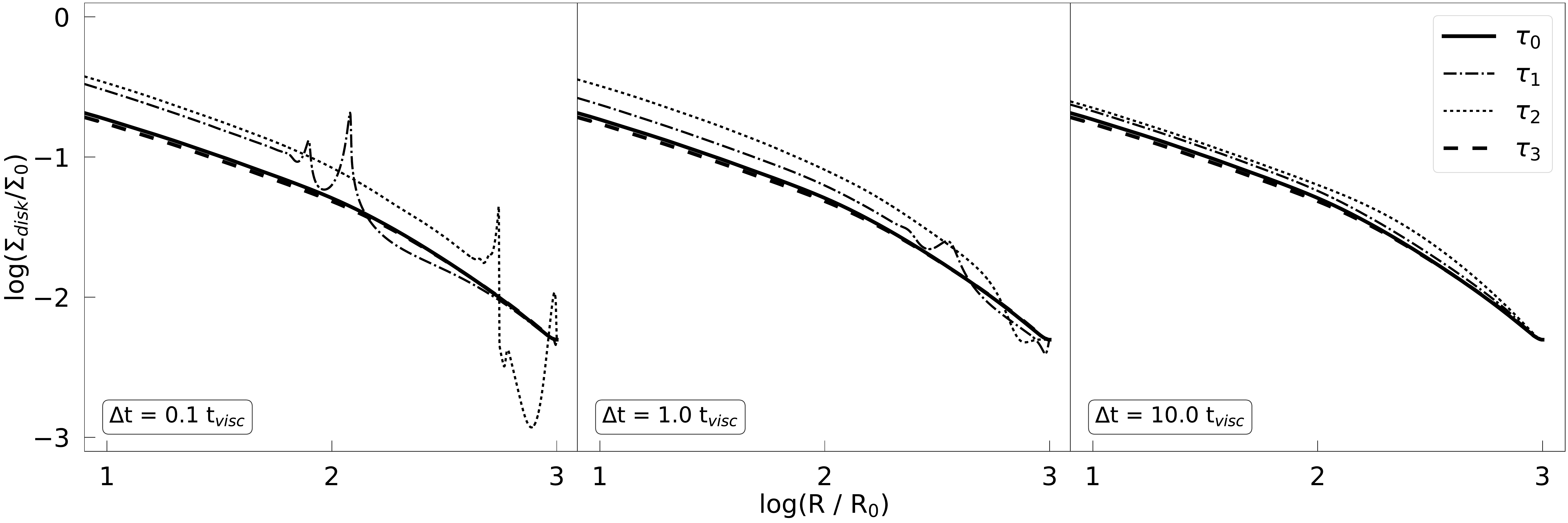}
        \caption{Simulation of the protoplanetary disk subjected to a density wave for different values $\Delta t = 0.1, 1.0, 10.0 \, t_\mathrm{visc}$ as the timescale for the gradual increase of the stellar mass. The plots show five arbitrary times starting from the initial stationary model ($\tau_0$; solid) across the transmission of the density wave ($\tau_1$ $\tau_2$; dash-dotted, dotted) until stationarity is attained again ($\tau_4$; dashed). The graphs in the panels only show about 30\% of the outer disk as here the density waves appear with the most intensity.}
        \label{f.density_waves_density2}
\end{figure}

As mass is transported inwards due to the gravitational pull of the host star and viscous friction within the disk (compare Sect.~\ref{s.phys_equations}), mass flow through the inner boundary of the disk is generated. Part of this mass is accreted onto the central star and therefore increasing the star's mass results in a change of the gravitational star-disk potential. A change in the gravitational potential directly results in an alteration of the Keplerian velocities in the disk. Areas close to the star will be affected much quicker than regions further from the centre and thus mass will be redistributed in the disk. This redistribution of mass is expected to be visible as an outwards propagating `density wave'. To test that our model reproduces this behaviour, we start from a stationary viscous disk and multiply the stellar mass by an arbitrary factor $\Delta m$. All other stellar parameters remain constant. Note that the exact value of $\Delta m$ only alters the intensity of the wave and therefore all our calculations have been carried out with $\Delta m = 1.5$. The change in the star's mass does not just increase the gravitational potential, but also alters the effective viscosity (cf. Sect.~\ref{ss.vis}). We emphasize that in our model a change in stellar mass instantly changes the Keplerian velocity and gravitational pull in each grid cell. Thus, the disk experiences an increased mass flow in each cell which leads to mass-flux instabilities if the time scale of the mass accretion onto the star $\Delta t$ exceeds the local time scale of the viscous disk $\tau_\nu = r^2\nu^{-1}$. Therefore, the mass accretion onto the star must be distributed over a timescale that is shorter than the viscous timescale of at a reference radius $\Delta t < \tau_\nu(r_\mathrm{ref})$ (cf. Fig.~\ref{f.density_waves_density2}). The shortest viscous timescale appears to be at the position with the highest effective viscosity. We found that $\tau_\nu$ at the inner boundary is always close to $\tau_\mathrm{\nu;min}$ and thus we set our reference radius to $r_\mathrm{ref} = r_\mathrm{in}$.

\begin{figure}[]
        \centering\includegraphics[width=\textwidth]{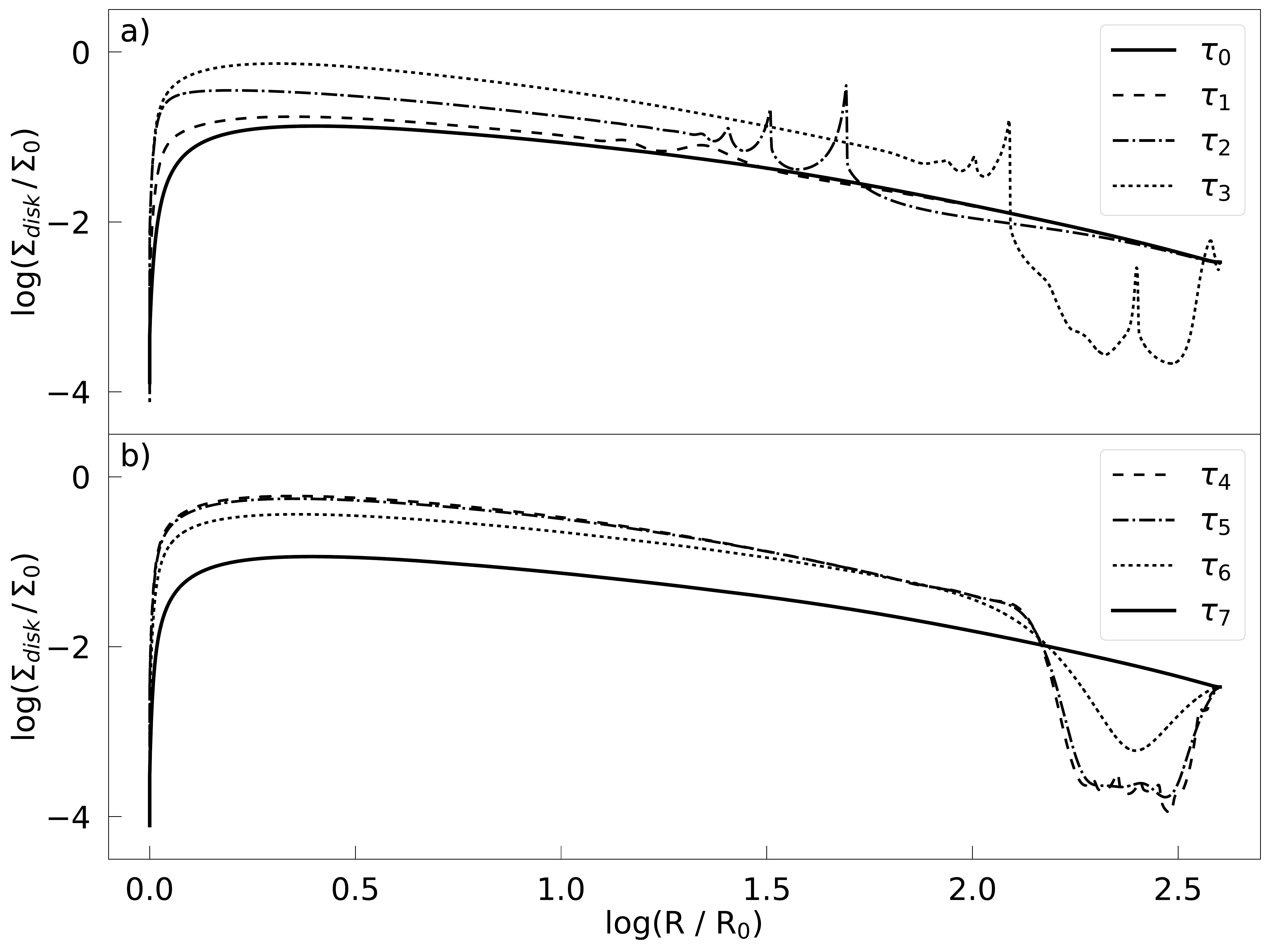}
        \caption{Propagation of the expected density wave in the density-radius space. All values are in dimensionless values. The black solid lines represent the initial (Panel a) and the final models (Panel b). Panel a shows the onset and propagation of the wave at several different times ($\tau_1$ to $\tau_3$, dashed). Once the wave reaches the outer boundary, the settling begins ($\tau_4$ to $\tau_6$, dot-dashed) which is shown in Panel b. The solution converges towards the final, stationary model ($\tau_7$) by slowly balancing the mass-flux at the inner and outer boundaries of the disk.}
        \label{f.density_waves_density}
\end{figure}

Figure~\ref{f.density_waves_density} shows the temporal evolution of the surface density after gradually increasing the stellar mass of a stationary viscous disk ($\tau_0$; solid, Panel a). As the central mass increases, mass accumulates at the inner boundary resulting in a higher surface density ($\tau_1$; dashed, Panel a). At disk regions further away from the centre, viscosity is not efficient enough to transport mass accordingly and a pile-up of mass occurs. This `mass-bumps' occur because density waves propagate towards the outer boundary ($\tau_2$ and $\tau_3$; dashed-dotted and dotted, Panel a). Once the density waves reach the outer boundary ($\tau_4$; dashed, Panel b), the mass flow within the disk settles but leaves the disk with an imbalanced mass accretion over the inner and outer boundaries (mass accretion is higher over the inner boundary due to more efficient viscosity). Hence, the disk loses mass ($\tau_5$ and $\tau_6$; dot-dashed and dotted, Panel b) until a stationary solution is reached. As expected, a higher central mass causes a larger gravitational pull and a higher rotational speed of the disk and thus, compared to the initial model, slightly less (in this case $\sim 1\%$) mass can be contained to maintain a stationary solution ($\tau_7$; solid, Panel b). These results show that our model can reproduce the behaviour expected in a scenario where the mass of the host star is increased and is therefore capable of producing comprehensive results when studying viscous disks.

We have mentioned that the mass accretion onto the central star has to be spread over a timescale that is shorter than the viscous timescale at a reference radius $r_\mathrm{ref}$. Since $r_\mathrm{ref}$ with the shortest viscous timescale is difficult to determine but the timescale on that mass is accreted onto the star might influence the result, we have studied the impact on the results when using different mass acrretion timescales. Figure~\ref{f.density_waves_density2} shows the behaviour of a stationary viscous disk when the central mass is increased by a factor of $\Delta m = 1.5$ (see above), with the mass accretion being distributed over three different timescales of $\Delta t = 0.1 \tau_\mathrm{\nu,in}$ (left panel), $\Delta t = 1 \tau_\mathrm{\nu,in}$ (middle panel) and $\Delta t = 10 \tau_\mathrm{\nu,in}$ (right panel), where $\tau_\mathrm{\nu,in}$ is the viscous time at the inner boundary. One can see that if mass is applied to the central star too slowly, the disk can adjust to the changing gravitational potential and the density waves get absorbed. This leads to the conclusion that if one wants to study mass infall correctly (e.g. when investigating FU-Ori bursts, infall of planets, etc.), it is even more important to compute the very inner disk regions consistently in order not to lose any physical feedback onto the disk. Moreover, density adjustments occur on timescales which are up to seven orders of magnitude larger than the mean Keplerian rotation period. Thus, for studying the long term evolution of astrophysical, viscous disks, only an implicit scheme can provide reasonably large timesteps. 

\subsection{Viscous evolution of the disk}
\label{ss.viscous_evol}
From the previous section, we have seen that viscosity plays a central role in the evolution of disks and thus the reliability of our viscosity model is an important factor.
Accretion disks can be considered as axisymmetric thin disks \cite{Armitage2013}. The disk can be separated into annuli with different average densities (the density distribution of one single annulus can be considered to be constant). According to $v_\mathrm{r} = r \Omega(r,t)$, the radial velocity is different for every disk radius and as the disk gas is a viscous fluid, friction between the annuli causes an energy loss from the innermost to the outermost disk regions. As a result, the inwards oriented mass-flux $\dot{m}$ occurs. If $\dot{m}$ is constant over all radial rings, the disk can be called a viscous stationary disk. If an external mass is applied anywhere onto the disk, the mass has to be dispersed over the disk (inwards and outwards) until the disk reaches a stationary solution again. To test our model, we consider an axisymmetric disk with a given surface density $\Sigma$. According to \cite{LyndenBell74} or \cite{Armitage2013}, we want to reproduce the time-dependent analytical solutions for the evolution of the surface density $\Sigma(r,t)$ of a geometrically thin disk under the action of internal angular momentum transport
\begin{equation}
    \frac{\partial \Sigma}{\partial t} = \frac{3}{r} \frac{\partial}{\partial r} \left[ \sqrt{r} \frac{\partial}{\partial r} \left( \nu \Sigma \sqrt{r}\right) \right] \:.
    \label{e.sigmadiff}
\end{equation}
This equation represents a diffusive partial differential equation for the surface density, which can be derived by applying a variable substitution, assuming constant viscosity $\nu = const.$. Defining
\begin{equation} \label{e.substitute}
    X = 2 \sqrt{r} \, \qquad\mbox{and}\qquad
    f = \frac{3}{2} \Sigma X \:,
\end{equation}
Equation~\ref{e.sigmadiff} takes the form of a typical diffusion equation
\begin{equation}
    \frac{\partial f}{\partial t} = D \frac{\partial^2 f}{\partial X^2} \, ,
\end{equation}
where $D$ is the diffusion coefficient given by
\begin{equation}
    D = \frac{12 \nu}{X^2} \:.
\end{equation}
Although a constant viscosity is not necessarily realistic for a protoplanetary disk, a Green's function solution and the qualitative illustration of the behaviour of Eq.~\ref{e.sigmadiff} is possible. Initial conditions at $t = 0$ for this problem are given in \cite{Armitage2013} where all mass is situated in a tenuous ring of mass $m$ at radius $r_0$
\begin{equation}
    \Sigma(r,t = 0) = \frac{m}{2 \pi r_0} \delta(r - r_0) \:,
    \label{e.dirac}
\end{equation}
where $\delta(r - r_0)$ is the Dirac delta function. Boundary conditions that enforce zero-torque at $r=0$ and free expansion at $r \rightarrow \infty$ yield (see \cite{LyndenBell74})
\begin{equation}
    \Sigma(x, \tau) = \frac{m}{\pi r_0^2} \frac{1}{\tau} x^{-\frac{1}{4}} \exp \left[ - \frac{(1+x^2)}{\tau} \right] I_{1/4} \left( \frac{2x}{\tau} \right)
    \label{e.sigarmitage}
\end{equation}
for the time-dependent solution for Eq.~\ref{e.sigmadiff}, where $x$ and $\tau$ represent unitless variables
\begin{subequations}
    \begin{align*}
        x & = \frac{r}{r_0} \\
        \tau & = 12 \nu r_0^{-2} t \:,
    \end{align*}
\end{subequations}
and $I_{1/4}$ is the modified Bessel function of the first kind.

Note that contrary to the pure analytical solution of \cite{Armitage2013}, our model solves a coupled system of differential equations and thus we have to adapt our numerical model to fit the analytical case. To do so, we start with a disk with an outer radius of $r_{out} = 10$~AU surrounding a central star with $M_\star = 1$~M$_\odot$. Our initial model has a density profile similar to Eq. \ref{e.dirac} and since a Dirac delta function is numerically not possible, we adopt a Gaussian distribution and normalized the surface density $\Sigma$ with an initially constant surface density $\Sigma_0$
\begin{equation}
    \Sigma(r) = \Sigma_0 \frac{r^2}{ \sqrt{ \pi r_0 } |a|} \exp \left[ - \frac{ \left( r - r_0 \right)^2}{a^2} \right] + \Sigma_{0} \:,
\end{equation}
where $\Sigma_\mathrm{0}$ represents a background density, $r_0 = \frac{1}{2} (r_\mathrm{out} - r_\mathrm{in})$ is the initial position of the grid cell and $a$ is comparable to the standard deviation. In our simulation, we solve the the equation of continuity, the equation of motion and the equation of energy for a constant kinematic viscosity $\nu$. Since our initial model needs time until it has adapted to the global viscosity and the density profile at $t = 0$ is not described by Eq. \ref{e.dirac}, we need a time correction factor $\Delta t$ to align with the analytical model. We found that if we substitute the averaged viscous time $\bar{t}_\mathrm{visc}$ for $\tau$ in Eq.~\ref{e.sigarmitage} so that $\tau_\mathrm{visc} = 12 \nu r_0^{-2} \bar{t}_\mathrm{visc}$, we can compute the $\Delta t$ that best fits numerical model best to the analytical model. The constant time correction factor is given by
\begin{equation}
    \Delta t = \frac{\bar{t}_{\nu}}{t} = \mathrm{const.}
\end{equation}
The remaining value $m$ in Eq.~ \ref{e.sigarmitage} is given by
\begin{equation}
    m = \sum 2 \pi r \Sigma(r) \Delta r \:.
\end{equation}

\begin{figure}[]
    \centering\includegraphics[width=\textwidth]{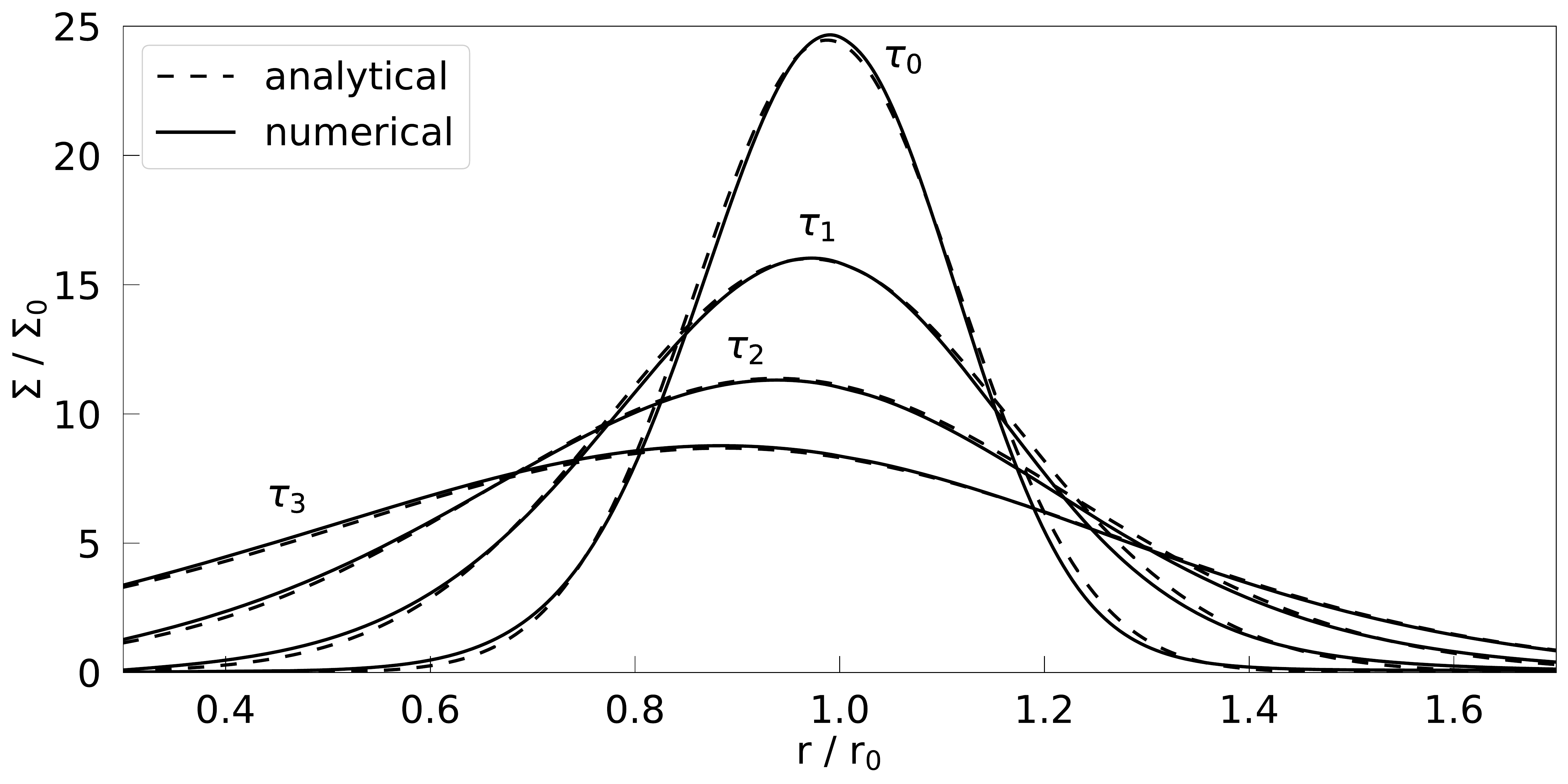}
    \caption{Comparison of the evolution of a surface density feature in a viscous disk with the analytical model presented by \cite{Armitage2013}. Numerical (solid lines) and analytical models (dashed lines) are plotted at different times ($\tau_0$ to $\tau_3$) in the surface density-radius space in unitless variables. We see that our viscosity model well reproduces the analytical solution and thus supports the correct implementation into our code.}
    \label{f.armitage}
\end{figure}

In Figure~\ref{f.armitage} we show the viscous evolution of a disk fragment with a background surface density profile $\Sigma_0$ on which we imposed a disturbance $\Sigma$ at a reference radius $r_0$, compared to the analytical solution at different times $\tau_\mathrm{n}$. The numerical results are represented by solid lines and the analytical results by dashed lines. One can see that with time the density feature is dissolving to the left and the right of $r_0$. Simultaneously, the maximum drifts inwards caused by the inwards-oriented mass-flux $\dot{m}$. Our model shows the same behaviour as predicted by the analytical solution and thus verifies the correctness of our viscosity model.

\subsection{Disk potential variation}
\label{ss.disk_potential_variation}
\begin{figure}[]
    \centering
        \includegraphics[width=0.9\textwidth]{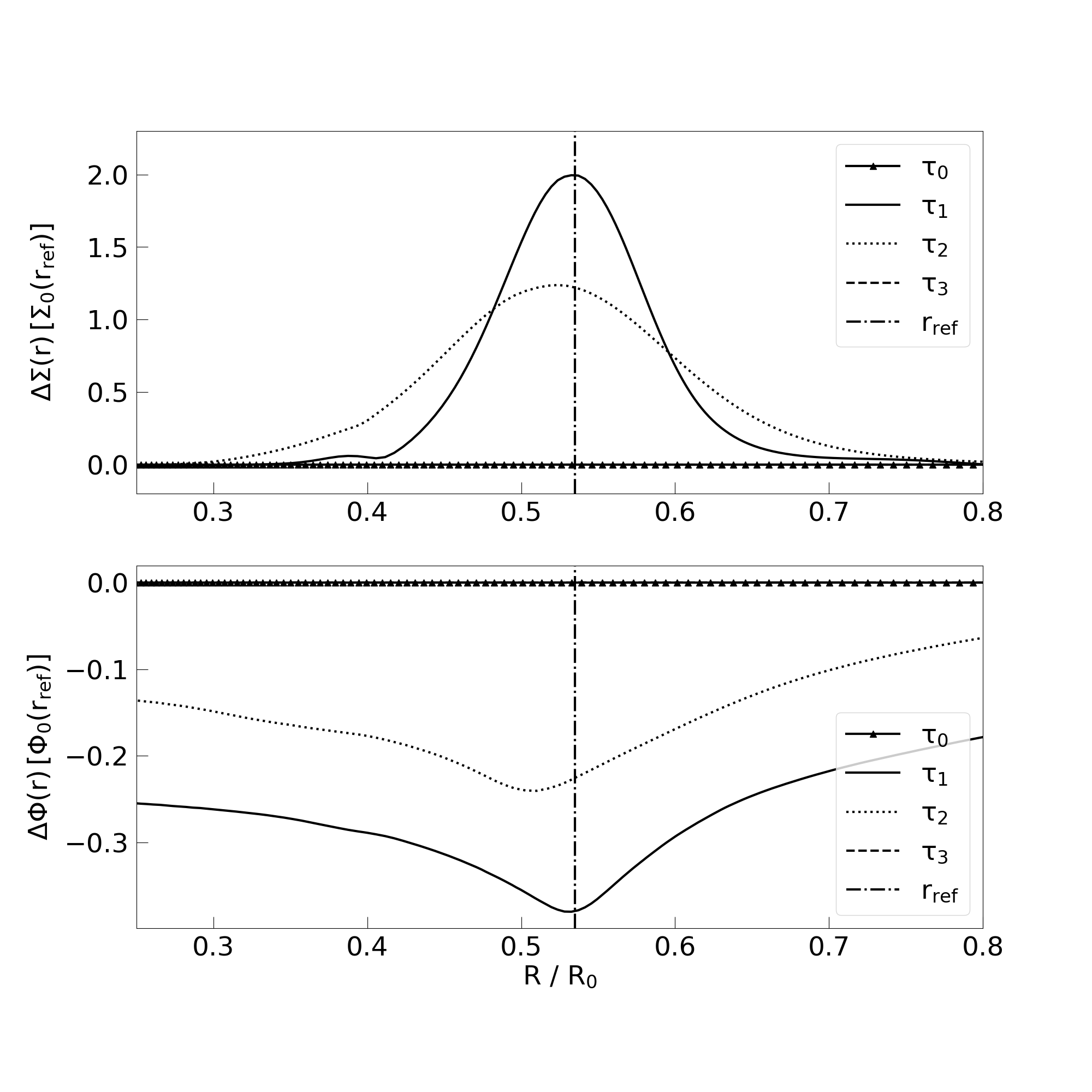}
    \caption{Reaction of the disk by perturbing the initial model with a mass $M_\mathrm{test}$ at an arbitrary radius $r_\mathrm{ref}$. The upper panel shows the evolution of the disk's relative surface density, $\Delta \Sigma(r)$, and the lower panel the adaptation of the relative gravitational potential, $\Delta \Psi(r)$, where $\Delta \Sigma(r) = \Sigma(r) - \Sigma_0(r_\mathrm{ref})$ and $\Delta \Psi(r) = \Psi(r) - \Psi_0(r_\mathrm{ref})$ respectively. The dash-dotted vertical line in both panels represents the position of the mass insertion $r_\mathrm{ref}$. On the horizontal axis, we have the radius in values of the reference radius $R_0 = R_{\rm out}$, where $R_{\rm out}$ represents the outer disk radius. Focusing on the lower panel, we see the formation of a potential cavity ($\tau_1$; solid) after applying the $M_\mathrm{test}$ to the initial model ($\tau_0$; solid-triangle). Due to viscous forces, mass will be transported inwards (to the left of $r_{\rm ref}$) and outwards (to the right of $r_{\rm ref}$), leading to a widening of the cavity ($\tau_2$; dotted). The drift of the cavity minimum to the left is caused by the inwards-oriented mass-flux. The mass distribution continues until the perturbation $M_{\rm test}$ is dispersed over the entire disk. This results in different mass accretion rates $\dot{m}$ over the inner and outer boundaries and consequently, due to the inwards-oriented mass-flux, to a mass loss over the inner boundary. The mass loss continues until the disk mass is equal to that of the initial model ($\tau_3$; dashed).}
    \label{f.disk_pot_mass1}
\end{figure}

In Section~\ref{ss.density_waves} we show that altering the gravitational potential of the star can lead to serious changes in the disk structure. Taking into account that our main goal is to investigate the behaviour of viscous disks, the transport of mass through the disk will affect the local viscosity due to changes in the disk's gravitational potential. Even though the gravitational potential of the disk is negligible compared to the star's potential, even small local changes might affect the global outcome when investigating the long term evolution of viscous disks.  
To test the correct implementation and behaviour of Eq.~\ref{e.disk_pot} and Eq.~\ref{e.disk_pot_acc}, we test whether the total gravitational potential is conserved, meaning $\int_{0}^{\infty} \Psi_{tot}(t) \, dr = \text{const.} \, \forall t$. As our computational domain is radially restricted from a few stellar-radii to several hundred AU, there is a difference between the fully analytical solution and the numerical result. However, the conservation-error in our simulations is less than $0.3$\% which is acceptable in the scope of astrophysical disk simulations. Furthermore, we emphasize that the disk's potential has to adapt to a change in the mass distribution of the disk. Hence, we tested this behaviour by inserting a mass $M_\mathrm{test}$, which is distributed over an annulus within the axisymmetric description at an arbitrary position $R_\mathrm{M}$ into the disk. Figure~\ref{f.disk_pot_mass1} (lower panel) shows the evolution of the disk potential at different times $\tau_\mathrm{i}$. The upper panel represents the simultaneous adaptation of the surface density. Focusing on the lower panel, we see the formation of a potential cavity ($\tau_1$, solid) after inserting $M_\mathrm{test}$ into the initial model ($\tau_\mathrm{0}$, solid-triangle). Due to viscous forces, $M_\mathrm{test}$ is dissolved radially ($\tau_2$, dotted), which results in a depletion of the cavity. The slightly higher disk-to-star mass ratio causes a change of accretion rate at the outer and inner boundary ($\dot{m}_\mathrm{in}$ and $\dot{m}_\mathrm{out}$). As this change in $\dot{m}$ is not equal for the inner and outer boundary (the inner region is affected more due to the gravitational pull from the host star), more mass is accreted over the inner boundary than the outer boundary, resulting in a mass loss of the disk. Once the disk mass is equivalent to the mass of the initial model, $\dot{m}$ is again constant over the entire disk, representing a stationary solution ($\tau_\mathrm{3}$; dashed) which is equivalent to the initial model. We emphasize that there is a maximum disk-to-star mass ratio if when exceeded, no stationary solution is possible \cite{Pringle1981}. A detailed investigation of this result should be considered in further disk evolution studies.

\section{Conclusion}
\label{s.conclusion}
In this paper, we present an implicit numerical method to solve axially symmetric time-dependent equations of radiation hydrodynamics. We assume hydrostatic equilibrium perpendicular to the equatorial plane, which can be described as a 1+1D approach. Although the numerical method is currently restricted to 1-dimensional problems, i.e.~all variables $X=X(r,t)$ depend on radius and time, such implicit computations allow an accurate description of axial symmetric configurations and their evolution over time scales much larger than several orbital periods. Global phenomena like mass accretion and angular momentum transport can be studied on long evolutionary timescales. In order to test the validity of our method, we have tested our model for various configurations.

Since the paper is focused on the numerical method, the presented examples are calculated with the simplest material functions, e.g.~an ideal equation of state, constant opacity, and constant viscosity. In our astrophysical computations, additional features are introduced by the more realistic material functions, i.e.~opacity changes due to ionization or dissociation or variations in the viscosity caused by the onset of magnetic instabilities \cite[e.g.]{Bai2014}. From spherical computations of pulsating stars \cite[e.g]{Dorfi2000}, we can conclude that such features can be nicely resolved by the adaptive grid and do not lead to additional complications as long as the derivatives on the dependent variables remain smooth.

We emphasize that radiation transport is an essential process for the thermal structure of the disk and thus has to be considered in astrophysical disk simulations. Changes in the thermal profile lead to changes in viscosity and consequently to a different mass flux through the disk. Since the radial radiation flux is only defined locally, because the disk is optically thick in radial direction, the vertical radiation transport becomes the dominant heating/cooling term. Since we adopt cylindrical geometry, a proper description for the radiation transport in vertical direction (see Sect.~\ref{s.heating_cooling}) has to be defined.

Furthermore, our results from Sect.~\ref{ss.bc} show that a consistent description of the entire disk is necessary in order to produce physically realistic results. Especially when leaving out the very inner parts of the disk, the structure of the disk changes drastically. Because of large orbital velocities in the central disk regions, explicit schemes are limited by the CFL condition and thus utilizing an implicit scheme can overcome this timestep restriction and produce more realistic long term evolution results.



\bibliographystyle{elsarticle-num}
\bibliography{main}



\end{document}